\documentclass[twocol]{ametsocV6.1}

\title{A Machine Learning Tutorial for Operational Meteorology, Part I: Traditional Machine Learning}
\authors{Randy J. Chase\correspondingauthor{Randy J. Chase, randychase@ou.edu}\aff{a,b,c}, David R. Harrison\aff{b,d,e}, Amanda Burke\aff{b,c}, Gary M. Lackmann\aff{f} and Amy McGovern\aff{a,b,c}} 

\affiliation{\aff{a}{School of Computer Science, University of Oklahoma, Norman OK USA}\\
\aff{b}{School of Meteorology, University of Oklahoma, Norman OK USA}\\
\aff{c}{NSF AI Institute for Research on Trustworthy AI in Weather, Climate, and Coastal Oceanography, University of Oklahoma, Norman OK USA}\\
\aff{d}{Cooperative Institute for Severe and High-Impact Weather Research and Operations, University of Oklahoma, Norman OK USA}\\
\aff{e}{NOAA/NWS/Storm Prediction Center, Norman, Oklahoma}\\
\aff{f}{Department of Marine, Earth, and Atmospheric Sciences, North Carolina State University, Raleigh, North Carolina}\\}

\abstract{Recently, the use of machine learning in meteorology has increased greatly. While many machine learning methods are not new, university classes on machine learning are largely unavailable to meteorology students and are not required to become a meteorologist. The lack of formal instruction has contributed to perception that machine learning methods are 'black boxes' and thus end-users are hesitant to apply the machine learning methods in their every day workflow. To reduce the opaqueness of machine learning methods and lower hesitancy towards machine learning in meteorology, this paper provides a survey of some of the most common machine learning methods. A familiar meteorological example is used to contextualize the machine learning methods while also discussing machine learning topics using plain language. The following machine learning methods are demonstrated: linear regression; logistic regression; decision trees; random forest; gradient boosted decision trees; naïve Bayes; and support vector machines. Beyond discussing the different methods, the paper also contains discussions on the general machine learning process as well as best practices to enable readers to apply machine learning to their own datasets. Furthermore, all code (in the form of Jupyter notebooks and Google Colaboratory notebooks) used to make the examples in the paper is provided in an effort to catalyse the use of machine learning in meteorology.}

\begin{document}

\maketitle

%
%
%
%
%
%

%

\section{Introduction}
\begin{figure*}[t]
 \centering
 \noindent\includegraphics[width=6in]{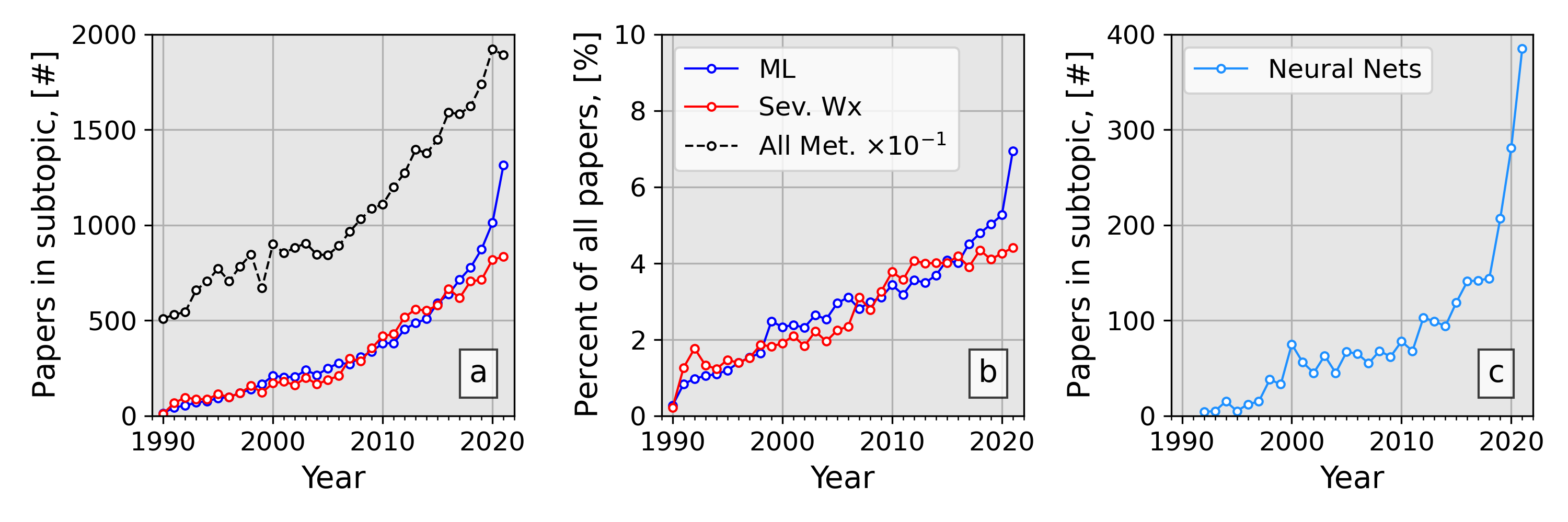}\\
 \caption{Search results for the Meteorology and Atmospheric Science category when searching abstracts for machine learning methods and severe weather. Machine learning keywords searched were: linear regression, logistic regression, decision trees, random forest, gradient boosted trees, support vector machines, k-means, k-nearest, empirical orthogonal functions, principal component analysis, self organizing maps, neural networks, convolutional neural networks and unets. Severe weather keywords searched were: tornadoes, hail, hurricanes and tropical cyclones. (a) Counts of publications per year for all papers in the Meteorology and Atmospheric Science category (black line; reduced by one order of magnitude), machine learning topics (blue line) and severe weather topics (red line). (b) Same as (a), but with the two subtopics normalized by the total number of Meteorology and Atmospheric Science papers. (c) Number of neural network papers (including convolutional and unets) published in Meteorology and Atmospheric sciences. All data are derived from Clarivate Web of Science.}\label{ml_history_pubs}
\end{figure*}

The mention and use of machine learning (ML) within meteorological journal articles is accelerating \citep[Fig. \ref{ml_history_pubs}; e.g.,][]{Burke2020,Hill2020,Lagerquist2020,Li2020,Loken2020,Mao2020,Munoz2020,Wang2020,Bonavita2021,Cui2021,Flora2021,Hill2021,Schumacher2021,Yang2021,Zhang2021}. With a growing number of published meteorological studies using ML methods, it is increasingly important for meteorologists to be well-versed in ML. However, the availability of meteorology specific resources about ML terms and methods is scarce. Thus, this series of papers (total of 2) aim to reduce the scarcity of meteorology specific ML resources. 

While many ML methods are generally not new (i.e., published before 2002), there is a concern from ML developers that end users (i.e., non-ML specialists) may be hesitant or are concerned about trusting ML. However, early work in this space suggests that non-technical explanations may be an important part of how end users perceive the trustworthiness of ML guidance \citep[e.g.,][]{Cains2022}. Thus, an additional goal of these papers is to enhance trustworthiness of ML methods through plain language discussions and meteorological examples. 

In practice, ML models are often viewed as a \textit{black box} which could also be contributing to user hesitancy. These mystified feelings towards ML methods can lead to an inherent distrust with ML methods, despite their potential. Furthermore, the seemingly opaque nature of ML methods prevents ML forecasts from meeting one of the three requirements of a good forecast outlined by \citet{Murphy1993}: consistency. In short, \citet{Murphy1993} explains that in order for a forecast to be good, the forecast must (1) be consistent with the user's prior knowledge, (2) have good quality (i.e., accuracy) and (3) be valuable (i.e., provide benefit). Plenty of technical papers demonstrate how ML forecasts can meet requirements 2 and 3, but as noted above if the ML methods are confusing and enigmatic, then it is difficult for ML forecasts to be consistent with a meteorologists prior knowledge. This series of papers will serve as a reference for meteorologists in order to make the \textit{black box} of ML more transparent and enhance user trust in ML.  

This paper is organized as follows. Section 2 provides an introduction to all ML methods discussed in this paper and will define common ML terms. Section 3 discusses the general ML methods in context of a simple meteorological example, while also describing the end-to-end ML pipeline. Then, Section 4 summarizes this paper and also discusses the topics of the next paper in the series. 

\section{Machine Learning Methods and Common Terms\label{sec:MethodsTerms}}

This section will describe a handful of the most common ML methods. Before that, it is helpful to define some terminology used within ML. First, we define ML as any empirical\footnote{By empirical we mean any method that uses data as opposed to physics} method where parameters are \textit{fit} (i.e., learned) on a training dataset in order to \textit{optimize} (e.g., minimize or maximize) a predefined \textit{loss} (i.e., cost) function. Within this general framework, ML has two categories: \textit{supervised} and \textit{unsupervised} learning. Supervised learning are ML methods that are trained with prescribed input \textit{features} and output \textit{labels}. For example, predicting tomorrow's high temperature at a specific location where we have measurements (i.e., labels). Meanwhile, unsupervised methods do not have a predefined output label \citep[e.g., self-organizing maps;][]{Nowotarski2013}. An example of an unsupervised ML task would be clustering all 500 mb geopotential height maps to look for unspecified patterns in the weather. This paper focuses on supervised learning.

The input features for supervised learning, also referred to as input data, predictors or variables, can be written mathematically as the vector (matrix) $X$. The desired output of the ML model is usually called the target, predictand or label, and is mathematically written as the scalar (vector) $y$. Drawing on the meteorological example of predicting tomorrow's high temperature, the input feature would be tomorrow's forecasted temperature from a numerical weather model (e.g., GFS) and the label would be tomorrow's observed temperature. 

Supervised ML methods can be further broken into two sub-categories: \textit{regression} and \textit{classification}. Regression tasks are ML methods that output a continuous range of values, like the forecast of tomorrow's high temperature (e.g., 75.0$^{\circ}$F). Meanwhile classification tasks are characteristic of ML methods that classify data (e.g., will it rain or snow tomorrow). Reposing tomorrow's high temperature forecast as a classification task would be: "Will tomorrow be warmer than today?". This paper will cover both regression and classification methods. In fact, many ML methods can be used for both tasks.

All ML methods described here will have one thing in common: the ML method quantitatively uses the training data to optimize a set of weights (i.e., thresholds) that enable the prediction. These weights are determined either by minimizing the error of the ML prediction or maximizing a probability of a class label. The two different methods coincide with the regression and classification respectively. Alternative names for error that readers might encounter in the literature are \textit{loss} or \textit{cost}.

Now that some of the common ML terms has been discussed, the following subsections will describe the ML methods. It will start with the simplest methods (e.g., linear regression) and move to more complex methods (e.g., support vector machines) as the sections proceed. Please note that the following subsections aim to provide an introduction and the intuition behind each method. An example of the methods being applied and helpful application discussion can be found in Section 3. 

\subsection{Linear Regression \label{sec:linear_regression}} 

\begin{figure}[t]
 \centering
 \noindent\includegraphics[width=2.5in]{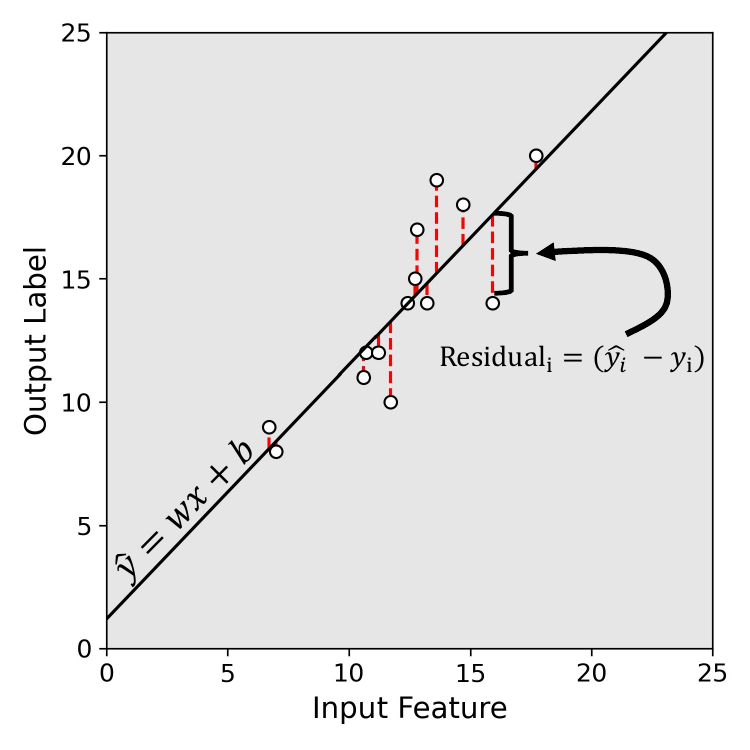}\\
 \caption{A visual example of linear regression with a single input predictor. The x-axis is a synthetic input feature, the y-axis is a synthetic output label. The solid black line is the regression fit, and the red dashed lines are the residuals.}\label{lin_regression}
\end{figure}

An important concept in ML is when choosing to use ML for a task, one should start with the simpler ML models first. Occam's razor\footnote{\url{https://en.wikipedia.org/wiki/Occam\%27s_razor}} tells us to prefer the simplest solution that can solve the task or represent the data. While this doesn't always mean the simplest ML model available, it does mean that simpler models should be tried before more complicated ones \citep{Holte:simple}. Thus, the first ML method discussed is linear regression which has a long history in meteorology \citep[e.g.][]{Malone1955} and forms the heart of the model output statistics product \citep[i.e., MOS;][]{Glahn1972} that many meteorologists are familiar with. Linear regression is popular because it is a simple method that is also computationally efficient. At its simplest form, linear regression approximates the value you would like to predict ($\hat{y}$) by fitting weight terms ($w_i$) in the following equation,
\begin{equation}
    \hat{y} = \sum_{i=0}^{i=D} w_i x_i .\label{e1}
\end{equation}
The first predictor ($x_0$) is always $1$ so that $w_0$ is a bias term, allowing the function to move from the origin as needed. $D$ is the number of features for the task. 

As noted before, with ML, the objective is to find $w_i$ such that a user-specified loss function (i.e., error function) is minimized. The most common loss function for traditional linear regression is the residual summed squared error (RSS):
\begin{equation}
    RSS = \sum_{j=1}^{N} (y_j - \hat{y}_j)^2 \label{e2}
\end{equation}
where $y_j$ is a true data point, $\hat{y}_j$ is the predicted data point and $N$ is the total number of data points in the training dataset. A graphical example of a linear regression and its residuals is shown in Fig. \ref{lin_regression}. Linear regression using residual summed squared error can work very well and is a fast learning algorithm, so we suggest it as a baseline method before choosing more complicated methods. The exact minimization method is beyond the scope of this paper, but know that the minimization uses the slope (i.e., derivative) of the loss function to determine how to adjust the trainable weights. If this sounds familiar, that is because it is the same minimization technique learned in most first year college calculus classes and is a similar technique to what is used in data assimilation for numerical weather prediction \citep[c.f., Chapter 5 and Section 10.5 in][]{kalnay_2002,lackmann_2011}. The concept of using the derivative to find the minimum is repeated throughout most ML methods given there is often a minimization (or maximization) objective.

Occasionally datasets can contain irrelevant or noisy predictors which can cause instabilities in the learning. One approach to address this is to use a modified version of linear regression known as ridge regression \citep{Hoerl1970}, which minimizes both the summed squared error (like before) and the sum of the squared weights called an $L_2$ penalty. Mathematically, the new loss function can be described as 
\begin{equation}
    RSS_{ridge} = \sum_{j=1}^{N} (y_j - \hat{y}_j)^2 + \lambda \sum_{i=0}^D w_i^2 \label{e3}
\end{equation}
Here, $\lambda$ (which is $\geq 0$) is a user-defined parameter that controls the weight of the penalty. Likewise, another modified version of linear regression is lasso regression \citep{Tibshirani1996} which minimizes the sum of the absolute value of the weights. This penalty to learning is also termed an $L_1$ penalty. The lasso loss function mathematically is 
\begin{equation}
    RSS_{lasso} = \sum_{j=1}^{N} (y_j - \hat{y}_j)^2 + \lambda \sum_{i=0}^D |w_i| \label{e4}
\end{equation}
Both lasso and ridge encourage the learned weights to be small but in different ways. The two penalties are often combined to create the elastic-net penalty \citep{Zou2005}
\begin{equation}
    RSS_{elastic} = \sum_{j=1}^{N} (y_j - \hat{y}_j)^2 + \lambda \sum_{i=0}^D (\alpha w_i^2 + (1-\alpha)|w_i|). \label{e5}
\end{equation}
In general, the addition of components to the loss function, like described in Eq. \ref{e3}-\ref{e5}, is known as \textit{regularization} and is found in other ML methods. Some recent examples of papers using linear regression include subseasonal prediction of tropical cyclone parameters \citep{Lee2020}, relating mesocyclone characteristics to tornado intensity \citep{Sessa2020} and short term forecasting of tropical cyclone intensity \citep{Hu2020}.

\subsection{Logistic Regression \label{sec:logistic_regression}}

\begin{figure}[t]
 \centering
 \noindent\includegraphics[width=2.5in]{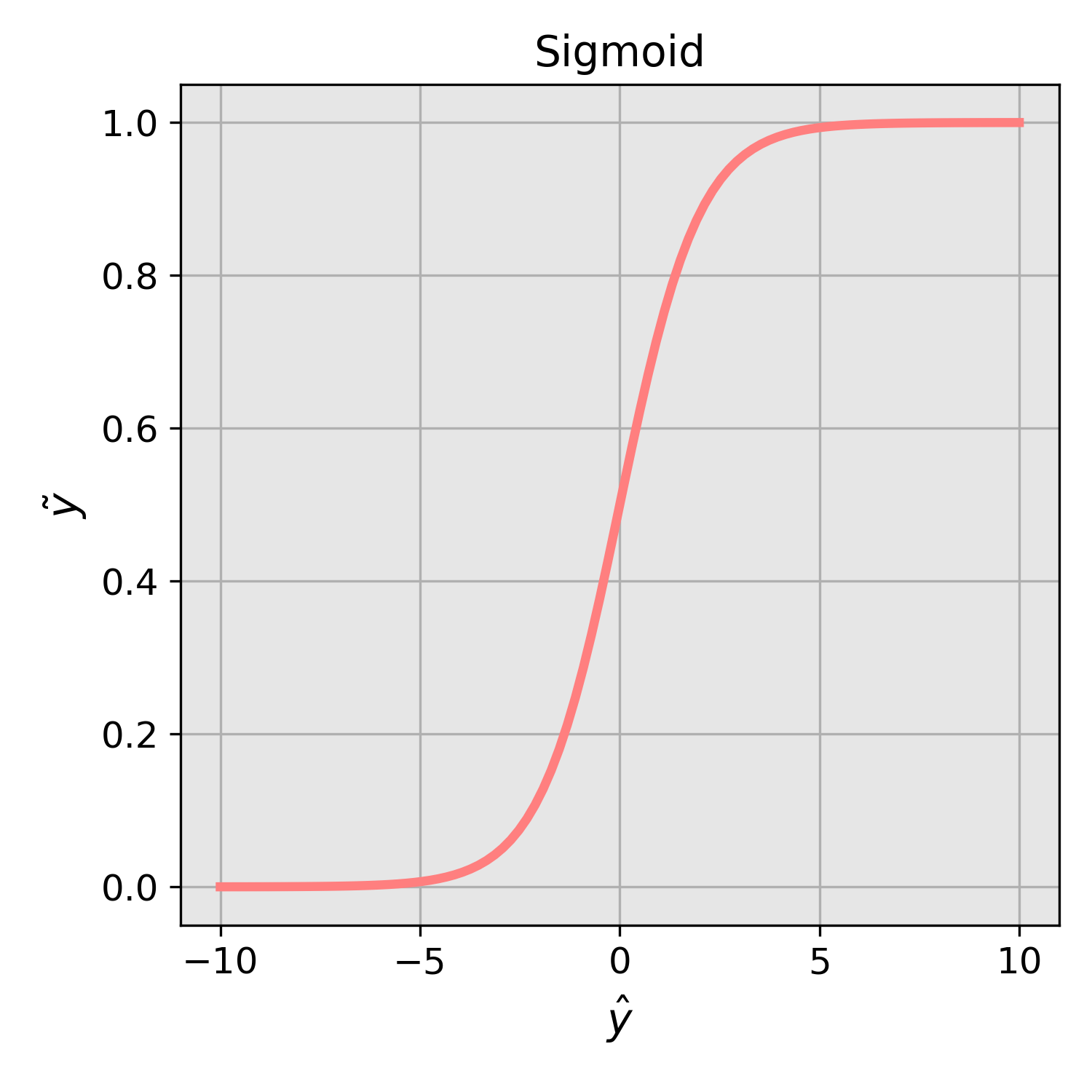}\\
 \caption{A graphical depiction of the sigmoid function (Eq. \ref{e6}). The x-axis is the predicted label value, while the y-axis is the now scaled value.}\label{sigmoid}
\end{figure}

As a complement to linear regression, the first classification method discussed here is logistic regression. Logistic regression is an extension from linear regression in that it uses the same functional form of Eq. \ref{e1}. The differences lie in how the weights for Eq. \ref{e1} are determined and a minor adjustment to the output of Eq. \ref{e1}. More specifically, logistic regression applies the sigmoid function (Fig. \ref{sigmoid}) to the output of Eq. \ref{e1} defined as: 
\begin{equation}
    S(\hat{y}) = \frac{1}{1+e^{-\hat{y}}} \label{e6}
\end{equation}
Large positive values into the sigmoid results in a value of 1 while large negative values result in a value of 0. Effectively, the sigmoid scales the output of Eq. \ref{e1} to a range of 0 to 1, which then can be interpreted like a probability. For the simplest case of classification involving just two classes (e.g., rain or snow), the output of the sigmoid can be interpreted as a probability of either class (e.g., rain or snow). The output probability then allows for the classification to be formulated as the $w_i$ that maximizes the probability of a desired class. Mathematically, the classification loss function for logistic regression can be described as 
\begin{equation}
    \mathrm{loss} =  \sum_{i=0}^{i=D} - y_i \log(S(\hat{y})) + (1-y_i) \log(1-S(\hat{y})). \label{e7}
\end{equation}
Like before for linear regression, the expression in Eq. \ref{e7} is minimized using derivatives. If the reader is interested in more information on the mathematical techniques of minimization they can find more information in Chapter 5 of \citet{kalnay_2002}. 

Logistic regression has been used for a long time within meteorology. One of the earliest papers using logistic regression showed skill in predicting the probability of hail greater than 1.9 cm \citep{Billet1997}, while more recent papers have used logistic regression to identify storm mode \citep{Jergensen2020}, subseasonal prediction of surface temperature \citep{Vigaud2019} and predict the transition of tropical cyclones to extratropical cyclones \citep{Bieli2020}. 
\subsection{Naïve Bayes \label{sec:NB}}
\begin{figure}[t]
 \centering
 \noindent\includegraphics[width=2.5in]{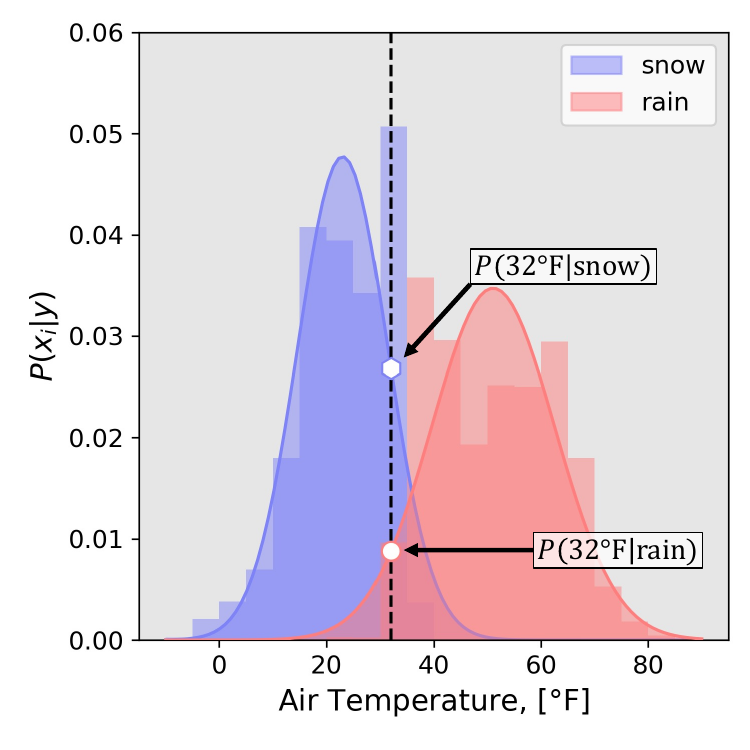}\\
 \caption{Visualizing the probability of a input feature given the class label. This example is created from five minute weather station observations from near Marquette, Michigan (years included: 2005 - 2020). Precipitation phase was determined by the present weather sensor. The histogram is the normalized number of observations in that temperature bin, while the smooth curves are the normal distribution fit to the data. Red are raining instances, blue are snowing instances.}\label{NB_Fig}
\end{figure}

An additional method to do classification is known as naïve Bayes \citep{Kuncheva2006}, which is named for its use of Bayes's theorem and can be written as the following:
\begin{equation}
    P(y|x)  = \frac{P(y)P(x|y)}{P(x)}. \label{e8}
\end{equation}
In words, Eq. \ref{e8} is looking for the probability of some label $y$ (e.g., snow), given a set of input features $x$ ($P(y|x)$; e.g., temperature). This probability can be calculated from knowing the probability of the label $y$ occurring in the dataset ($P(y)$; e.g., how frequent it snows) times the probability of the input features given it belongs to the class $y$ ($P(x|y)$; e.g., how frequently is it 32$^\mathrm{o}$F when it's snowing), divided by the probability of the input features ($P(x)$). The \textit{naïve} part of the naïve Bayes algorithm comes from assuming that all input features $x$, are independent of one another and the term $P(x|y)$ can be modeled by an assumed distribution (e.g., normal distribution) with parameters determined from the training data. While these assumptions are often not true, the naïve Bayes classifier can be skillful in practice. A few simplification steps results in the following
\begin{equation}
    \hat{y}  = \mathrm{argmax}(\log(P(y)) + \sum_{i=0}^{N}{\log(P(x_{i}|y))}). \label{e9}
\end{equation}
Again in words, the predicted class ($\hat{y}$) from naïve Bayes is the classification label ($y$) such that the sum of the log of the probability of that classification ($P(y)$) and the sum of log of all the probabilities of the specific inputs given the classification ($P(x_i|y)$) is maximized. In order to help visualize the quantity $P(x_i|y)$, a graphical example is shown in Fig. \ref{NB_Fig}. This example uses surface weather measurements from a station near Marquette, Michigan where data were compiled when it was raining and snowing. Fig. \ref{NB_Fig} shows distribution of air temperature (i.e., an input feature) given the two classes (i.e., rain vs snow). In order to get $P(x_i|y)$, we need to assume an underlying distribution function. The common assumed distribution with naïve Bayes is the normal distribution
\begin{equation}
    f(x; \mu, \sigma)  = \frac{1}{\sigma \sqrt{2\pi}} e^{-\frac{1}{2}(\frac{x-\mu}{\sigma})} \label{e10}
\end{equation}
where $\mu$ is the mean and $\sigma$ is the standard deviation of the training data. While the normal distribution assumption for the temperature distribution in Fig. \ref{NB_Fig} is questionable due to thermodynamic constraints that \textit{lock} the temperature at 32$^{\mathrm{o}}$F (i.e., latent cool/heating), naïve bayes can still have skill. Initially, it might not seem like any sort of weights/biases are being fit like the previously mentioned methods (e.g., logistic regression), but $\mu$ and $\sigma$ are being \textit{learned} fron the training data. If performance from the normal distribution is poor, other distributions can be assumed, like a multinomial or a Bernoulli distribution. 

A popular use of naïve Bayes classification in the meteorological literature has been the implementation of ProbSevere \citep[e.g.,][]{Cintineo2014,Cintineo2018,Cintineo2020} which uses various severe storm parameters and observations to classify the likelihood of any storm becoming severe in the next 60 minutes. Additional examples of naïve Bayes classifiers in meteorology have been used for identifying tropical cyclone secondary eyewall formation from microwave imagery \citep{Kossin2009}, identifying anomalous propagation in radar data \citep{Peter2013} and precipitation type (e.g., Convective/Stratiform) retrievals from geostationary satellites \citep{Grams2016}.

\subsection{Trees and Forests \label{sec:trees}}

\begin{figure}[t]
 \centering
 \noindent\includegraphics[width=2.5in]{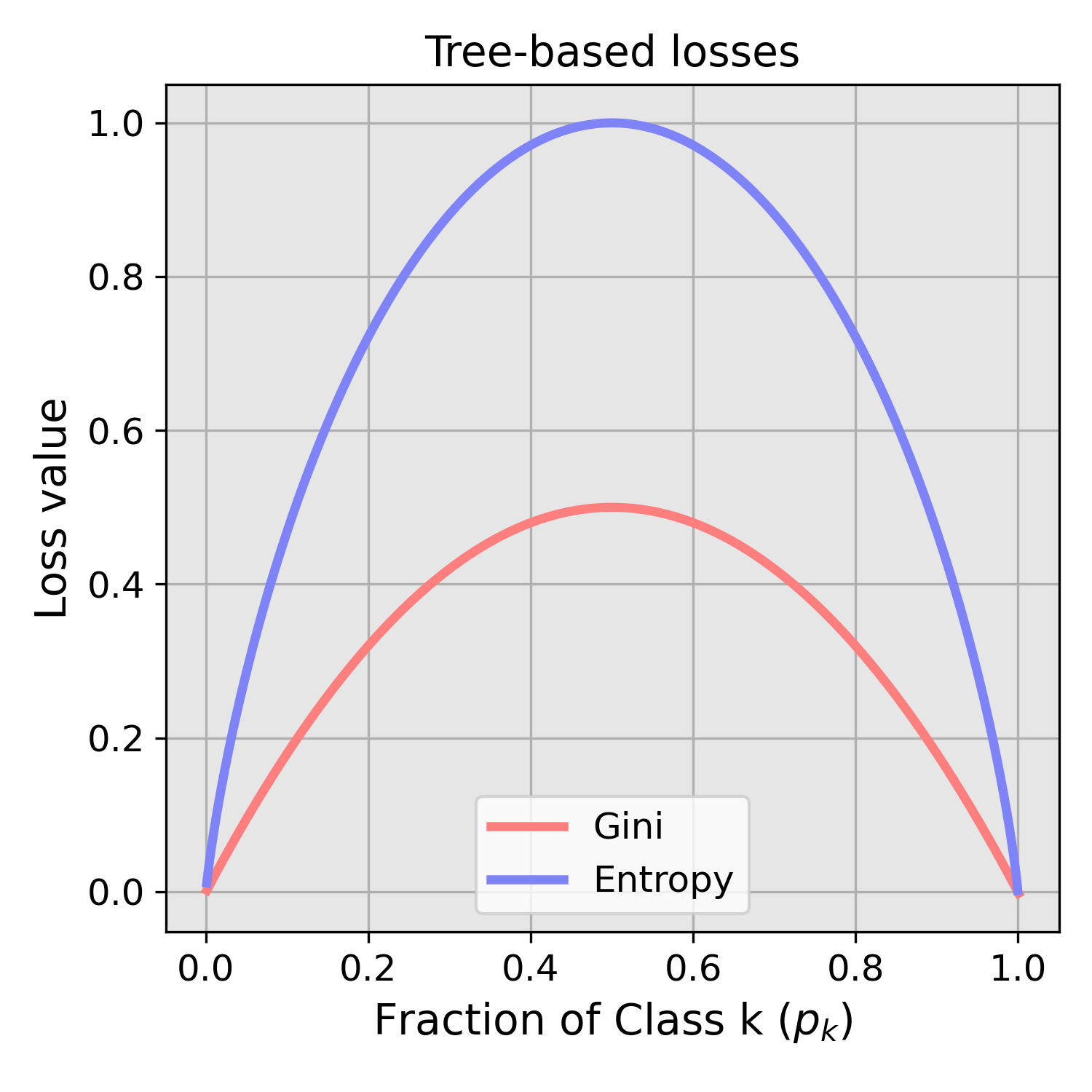}\\
 \caption{A visual representation of the two functions that can be used in decision trees for classification, Entropy (blue) and Gini impurity (red).}\label{tree_class_loss}
\end{figure}
Decision trees are based on a decision making method that humans have been using for years: flow charts, where the quantitative decision points within the flow chart are learned automatically from the data. Early use of decision trees in meteorology \citep[e.g.,][]{Chisholm1968} actually pre-dated the formal description of the decision tree algorithm \citep{Breiman1984,Quinlan1993,Breiman2001}. Since then, tree-based methods have grown in popularity and have been demonstrated to predict a variety of complex meteorological phenomena. Topics include: aviation applications \citep[e.g.,][]{Williams2008vil,Williams2008cit,Williams2014,Munoz2020}; severe weather \citep[e.g.,][]{Gagne2009,Gagne2013,McGovern2014enhancing,Mecikalski2015,Lagerquist2017,Gagne2017_hail,Czernecki2019,Burke2020,Hill2020,Loken2020,Gensini2021,Flora2021,Loken2022}; solar power \citep[e.g.,][]{McGovern2015solar}; precipitation \citep[e.g.,][]{ElmoreMpingRandomForests2016,Herman2018a,Herman2018b,Taillardat2019,Loken2020,Wang2020,Mao2020,Li2020,Hill2021,Schumacher2021}; satellite and radar retrievals \citep[e.g.,][]{Kuhnlein2014,Conrick2020,Yang2021,Zhang2021} and climate related topics \citep[e.g.,][]{Cui2021}.

To start, we will describe decision trees in context of a classification problem. The decision tree creates splits in the data (i.e., decisions) that are chosen such that either the Gini Impurity value or the Entropy value decreases after the split. Gini Impurity is defined as 
\begin{equation}
    \mathrm{Gini} = \sum_{i=0}^{i=k} p_i(1 - p_i) \label{e11}
\end{equation}
where $p_i$ is the probability of class i (i.e., the number of data points labeled class i divided by the total number of data points). While Entropy is defined as
\begin{equation}
    \mathrm{Entropy} = \sum_{i=0}^{i=k} p_i\log_2(p_i). \label{e12}
\end{equation}
Both functions effectively measure how similar the data point labels are in each one of the groupings of the tree after some split in the data. Envision the flow chart as a tree. The decision is where the tree branches into two directions, resulting in two separate leaves. The goal of a decision tree is to choose the branch that results in a leaf having a minimum of Gini or Entropy. In other words, the data split would ideally result in two sub-groups of data where all the labels are the same within each sub-group. Fig. \ref{tree_class_loss} shows both the Gini impurity and entropy for a two class problem. Consider the example of classifying winter precipitation as rain or snow. From some example surface temperature dataset the likely decision threshold would be near 32$^{\circ}$F, which would result in the subsequent two groupings of data point labels (i.e., snow/rain) having a dominant class label (i.e., fraction of class k is near 0 or 1) and thus having a minimum of Entropy or Gini (i.e., near 0). The actual output of this tree could be either the majority class label, or the ratio of the major class (i.e., a probabilistic output).

While it is helpful to consider a decision tree with a single decision, also known as a tree with a depth of 1, the prediction power of a single decision is limited. A step toward more complexity is to include increasing depth (i.e., more decisions/branches). To continue with the rain/snow example from the previous paragraph, we could include a second decision based on measured wet bulb temperature. A tree with depth two will likely have better performance, but the prediction power is still somewhat limited. 

An additional step to increase the complexity of decision trees, beyond including more predictors, is a commonly used method in meteorology: ensembles. While it might not be clear here, decision trees become over-fit (i.e., work really well for training data, but perform poorly on new data) as the depth of the tree increases. An alternative approach is to use an ensemble of trees (i.e., a forest). Using an ensemble of trees forms the basis of two additional tree based methods: random forests \citep[][]{Breiman2001} and gradient boosted decision trees \citep{Friedman2001}.

Random forests are a collection of decision trees that are trained on random subsets of data and random subsets of input variables from the initial training dataset. In other words, the mathematics are exactly the same for each tree, the decisions still aim to minimize the loss (e.g., Entropy), but each tree is given a different random subset of data sampled from the original dataset with replacement. Gradient boosted decision trees are an ensemble of trees that instead of training multiple trees on random subsets (i.e., random forest), each tree in the ensemble is successively trained on the remaining error from the previous trees. To put it another way, rather than minimizing the total error on random trees, the reduced error from the first decision tree is now minimized on the second tree, and the reduced error from trees one and two is then minimized on the third tree and so on. In order to come up with a single prediction out of the ensemble of trees, the predictions can be combined through a voting procedure (i.e., count up the predicted classes of each tree) or by taking the average probabilistic output from each tree. Random forests can use either method, while gradient boosted trees are limited to the voting procedure.

While the discussion here has been centered on classification for the tree-based methods, they can be used for regression as well. The main alteration to the decision tree method to convert to a regression-based problem is the substitution of the loss function (i.e., Eq. \ref{e11}-\ref{e12}). For example a common loss function for random forest for regression and gradient boosted regression is the same loss function as linear regression described in the previous section (e.g., Eq. \ref{e2}), the residual summed squared error.

\subsection{Support Vector Machines \label{sec:SVM}}

\begin{figure*}[t]
 \centering
 \noindent\includegraphics[width=6in]{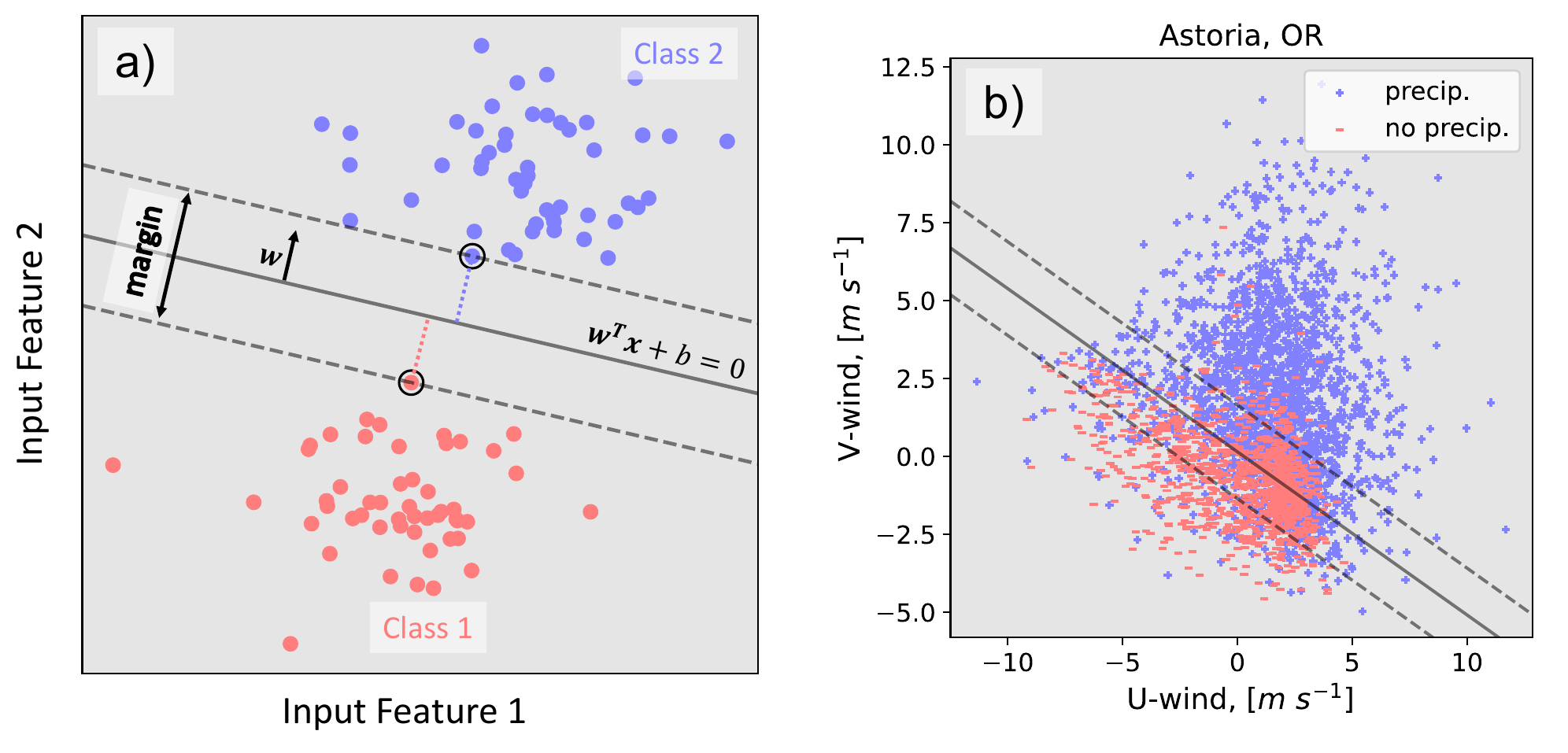}\\
 \caption{Support vector machine classification examples. (a) ideal (synthetic) data where the x and y axis are both input features, while the color designates what class each point belongs to. The decision boundary learned by the support vector machine is the solid black line, while the margin is shown by the dashed lines. (b) a real world example using NAM18Z forecasts of U and V wind and tipping bucket measurements of precipitation. Blue plus markers are raining instances and the red minus signs are non-raining instances. Black lines are the decision boundary and margins.}\label{SVM_Fig}
\end{figure*}

A support vector machine \citep[commonly referred to as SVM;][]{Vapnik1963} is an ML method similar to linear and logistic regression. The idea is that a support vector machine uses a linear boundary to do its predictions, which has a similar mathematical form but written differently to account to vector notation. The equation is 
\begin{equation}
    \hat{y} = \textbf{w}^{T}\textbf{x} + b \label{e13}
\end{equation}
where $\textbf{w}$ is a vector of weights, $\textbf{x}$ is a vector of input features, b is a bias term and $\hat{y}$ is the regression prediction. In the case of classification, only sign of the right side of Eq. \ref{e13} is used. This linear boundary can be generalized beyond two-dimensional problems (i.e., two input features) to three-dimensions where the decision boundary is called a plane, or any higher order space where the boundary is called a hyperplane. The main difference between linear methods discussed in Sections 2a-2b and support vector machines is that support vector machines include margins to the linear boundary. Formally, the margin is the area between the linear boundary and the closest training datapoint for each class label (e.g., closest rain data point and closest snow datapoint). This is shown schematically with a synthetic dataset in Fig. \ref{SVM_Fig}a. While this is an ideal case, usually classes overlap (Fig. \ref{SVM_Fig}b), but support vector machines can still handle splitting the classes. The optimization task for support vector machines is stated as the following: Find $\textbf{w}^{T}$ such that the margin is maximized. In other words, support vector machines aim to maximize the distance between the two closest observations on either side of the hyperplane. Mathematically, the margin distance is described as 
\begin{equation}
    \mathrm{margin} = \frac{1}{\textbf{w}^T \textbf{w}}. \label{e14}
\end{equation}
Like before, the maximization is handled by numerical techniques to optimize the problem but the resulting solution will be the hyperplane with the largest separation between the classes. A powerful attribute of the support vector machine method is that it can be extended to additional mathematical formulations for the boundary, for example a quadratic function. Thus the person using support vector machines can decide which function would work best for their data. Recent applications of support vector machines in meteorology include the classification of storm mode \citep{Jergensen2020}, hindcasts of tropical cyclones \citep{Neetu2020} and evaluating errors with quantitative precipitation retrievals in the United States \citep{Kurdzo2020}.


\section{Machine Learning Application and Discussion \label{sec:ml_app_dis}}

\begin{figure*}[t]
 \centering
 \noindent\includegraphics[width=5.5in]{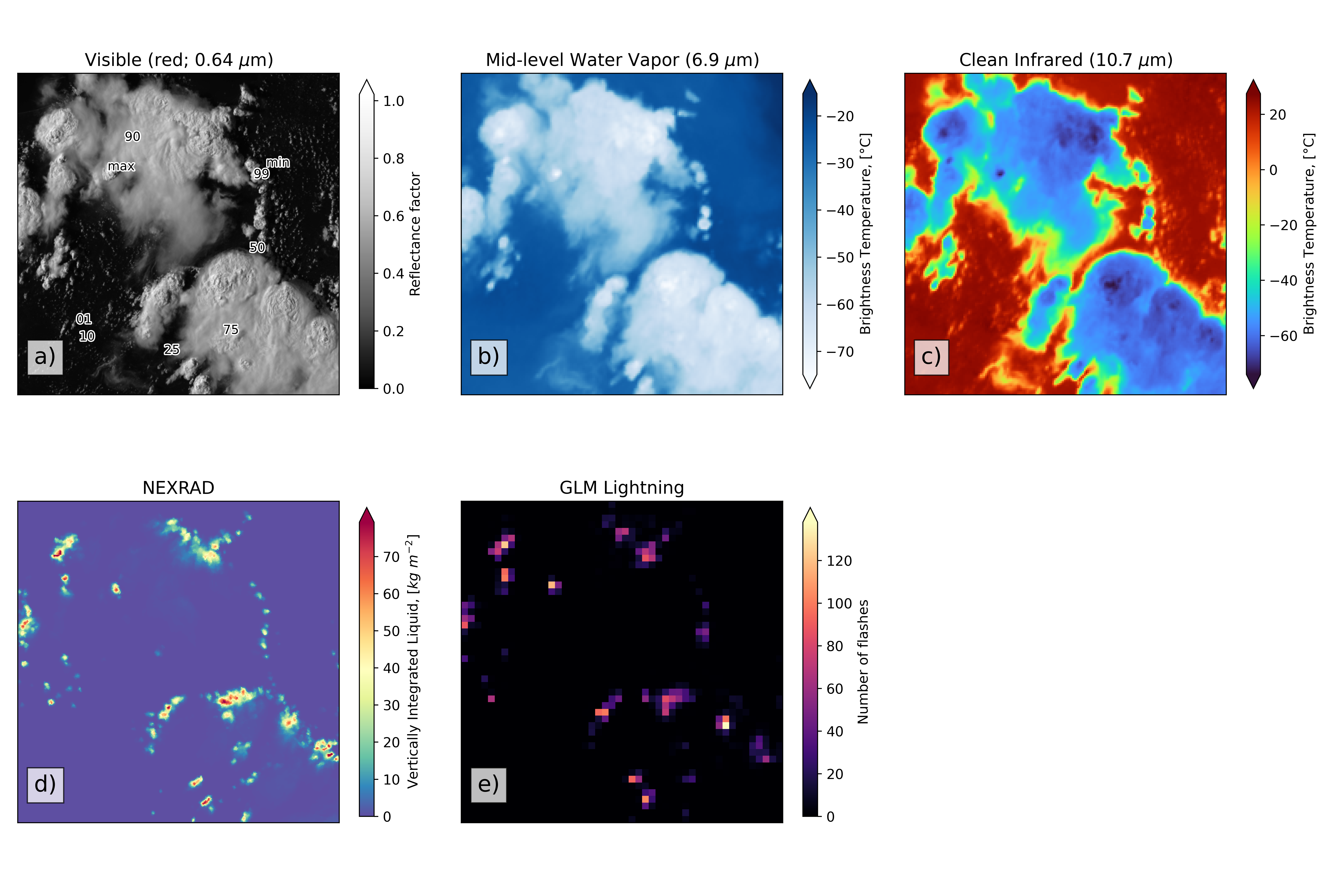}\\
 \caption{An example storm image from the Storm EVent ImageRy dataset. This event is from 06 August 2018. (a) the visible reflectance (b) the mid-tropospheric water vapor brightness temperature (c) the clean infrared brightness temperatures (d) the vertically integrated liquid retrieved from NEXRAD and (e) gridded GLM number of flashes. (a) also has annotated locations of representative percentiles that were engineered features used for the ML models.}\label{SEVIR_fig}
\end{figure*}

This section will discuss the use of all ML methods with a familiar use-case: thunderstorms. Specifically, this section will show two ML applications derived from popular meteorological datasets: radar and satellite. The particular data used are from the Storm EVent ImageRy dataset \citep[SEVIR;][]{Veillett2020}, which contains over 10,000 storm events from between 2017 and 2019. Each event spans four hours and includes measurements from both GOES-16 and NEXRAD. An example storm event and the 5 measured variables: Red channel visible reflectance (0.64$\mu$m; Channel 2), midtropospheric water vapor brightness temperature (6.9 $\mu$m; Channel 9), clean infrared window brightness temperature (10.7 $\mu$m; Channel 13), Vertically Integrated Liquid (VIL; from NEXRAD) and Geostationary Lightning Mapper (GLM) measured lightning flashes are found in Fig. \ref{SEVIR_fig}. In addition to discussing ML in context of the SEVIR dataset, this section will follow the general steps to using ML and contain helpful discussions of the best practices as well as the most common pitfalls.

\subsection{Problem Statements \label{sec:problem}}
The SEVIR data will be applied to two tasks: (1) Does this image contain a thunderstorm?  and (2) How many lightning flashes are in this image? To be explicit, we assume the GLM observations are unavailable and we need to use the other measurements (e.g., infrared brightness temperature) as features to estimate if there are lightning flashes (i.e., classification), and how many of them are there (i.e., regression). While both of these tasks might be considered redundant since we have GLM, the goal of this paper is to provide discussion on how to use ML as well as discussion on the ML methods themselves. That being said, a potential useful application of the trained models herein would be to use them on satellite sensors that do not have lightning measurements. For example, all generations of GOES prior to GOES-16 did not have a lightning sensor co-located with the main sensor. Thus, we could potentially use the ML models trained here to estimate GLM measurements prior to GOES-16 (i.e., November 2016).  

\subsection{Data \label{sec:data}}
The first step of any ML project is to obtain data. Here, the data are from a public archive hosted on the Amazon Web Service. For information of how to obtain the SEVIR data as well as the code associated with this manuscript see the \textit{Data Availability Statement}. One major question at this juncture is, "How much data is needed to do machine learning?". While there does not exist a generic number that can apply to all datasets, the idea is to obtain enough data such that one's training data are diverse. A diverse dataset is desired because any bias found within the training data would be encoded in the ML method \citep{McGovern2021}. For example, if a ML model was trained on only images where thunderstorms were present, then the ML model would likely not know what a non-lightning producing storm would look like and be biased. Diversity in the SEVIR dataset is created by including random images (i.e., no storms) from all around the United States \citep[c.f. Figure 2 in][]{Veillett2020}.

After obtaining the data, it is vital to remove as much spurious data as possible before training because the ML model will not know how to differentiate between spurious data and high quality data. A common anecdote when using ML models is \textit{garbage in, garbage out}. The SEVIR dataset has already gone through rigorous quality control, but this is often not the case with raw meteorological datasets. Two examples of quality issues that would likely be found in satellite and radar datasets are satellite artifacts \citep[e.g., GOES-17 heat pipe;][]{McCorkel2019} and radar ground clutter \citep[e.g.,][]{Hubbert2009}. Cleaning and manipulating the dataset to get it ready for ML often takes a researcher 50$\%$ - 80$\%$ of their time\footnote{\url{https://www.nytimes.com/2014/08/18/technology/for-big-data-scientists-hurdle-to-insights-is-janitor-work.html}}. Thus, do not be discouraged if cleaning one's datasets is taking a large amount of time because a high-quality dataset will be best for having a successful ML model. 

Subsequent to cleaning the data, the next step is to engineer the inputs (i.e., features) and outputs (i.e., labels). One avenue to create features is to use every single pixel in the image as a predictor. While this could work, given the number of pixels in the SEVIR images (589,824 total pixels for one visible image) it is computationally impractical to train a ML model with all pixels. Thus, we are looking for a set of statistics than can be extracted from each image. For the generation of features, domain knowledge is critical because choosing meteorologically relevant quantities will ultimately determine the ML models skill. For the ML tasks presented in Section 3a, information about the storm characteristics (e.g., strength) in the image would be beneficial features. For example, a more intense storm is often associated with more lightning. Proxies for estimating storm strength would be: the magnitude of reflectance in the visible channel; how cold brightness temperatures in the water vapor and clean infrared channel are; and how much vertically integrated water there is. Thus, to characterize these statistics, we extract the following percentiles from each image and variable: 0,1,10,25,50,75,90,99,100. 

To create the labels the number of lightning flashes in the image are summed. For Problem Statement 1, an image is classified as containing a thunderstorm if the image has at least one flash in the last five minutes. For Problem Statement 2, the sum of all lightning flashes in the past five minutes within the image are used for the regression target. 

Now that the data have been quality controlled and our features and labels have been extracted, the next step is to split that dataset into three independent sub-categories named the \textit{training}, \textit{validation} and \textit{testing} sets. The reason for these three sub-categories is because of the relative ease at which ML methods can "memorize" the training data. This occurs because ML models can contain numerous (e.g., hundreds, thousands, or even millions) learnable parameters, thus the ML model can learn to perform well on the training data but not generalize to other non-training data, which is called \textit{over-fitting}. In order to assess how over-fit a ML model is, it is important to evaluate a trained ML model on data outside of its training data (i.e., validation and testing sets).

The training dataset is the largest subset of the total amount of data. The reason the training set is the largest is because the aforementioned desired outcome of most ML models is to generalize on wide variety of examples. Typically, the amount of training data is between 70 - 85$\%$ of the total amount of data available. The validation dataset, regularly 5-15$\%$ of the total dataset, is a subset of data used to assess if a ML model is over-fit and is also used for evaluating best model configurations (e.g., the depth of a decision tree). These model configurations are also known as \textit{hyperparameters}. Machine learning models have numerous configurations and permutations that can be varied and could impact the skill of any one trained ML model. Thus, common practice is to systematically vary the available hyperparameter choices, also called a grid search, and then evaluate the different trained models based on the validation dataset. Hyperparamters will be discussed in more detail later. The test dataset is the last grouping that is set aside to the very end of the ML process. The test dataset is often of similar size to the validation dataset, but the key difference is that the test dataset is used \textbf{after} all hyperparameter variations have been concluded. The reason for this last dataset is because when doing the systematic varying of the hyperparameters the ML practitioner is inadvertently tuning a ML model to the validation dataset. One will often choose specific hyperparameters in such a way to achieve the best performance on the validation dataset. Thus, to provide a truly unbiased assessment of the trained ML model skill for unseen data, the test dataset is set aside and not used until after training all ML models. 

It is common practice outside of meteorology (i.e., data science) to randomly split the total dataset into the three subsets. However, it is important to strive for independence of the various subsets. A data point in the training set should not be highly correlated to a data point in the test set. In meteorology this level of independence is often challenging given the frequent spatial and temporal auto-correlations in meteorologic data. Consider the SEVIR dataset. Each storm event has four hours of data broken into five minute time steps. For one storm event, there is a large correlation between adjacent five minute samples. Thus, randomly splitting the data would likely provide a biased assessment of the true skill of the ML model. In order to reduce the number of correlated data points across subsets, time is often used to split the dataset. For our example, we choose to split the SEVIR data up by training on 01 Jan 2017 - 01 Jun 2019 and split every other week in the rest of 2019 into the validating and testing sets. This equates to a 72$\%$, 13$\%$ and 15$\%$ split for the training, validation and test sets respectively. In the event that the total dataset is small and splitting the data into smaller subsets creates less robust statistics, a resampling method known as \textit{k-fold cross-validation} \citep[e.g.,][]{Bischl2012,Goodfellow-et-al-2016} can be used. The SEVIR dataset was sufficiently large that we chose not to do k-fold cross-validation, but a meteorological example using it can be found in \citet{Shield2022}.



\subsection{Training and Evaluation \label{sec:train_n_eval}}

\subsubsection{Classification}
As stated in Section 3.a, task (1) is to classify if an image contains a thunderstorm. Thus, the classification methods available to do this task are: logistic regression, naïve Bayes, decision trees, random forest, gradient boosted trees and support vector machines. In order to find an optimal ML model, it is often best to try all methods available. While this might seem like a considerable amount of additional effort, the ML package used in this tutorial (i.e., scikit learn\footnote{\url{https://scikit-learn.org/stable/}}) uses the same syntax for all methods (e.g., method.fit($X$,$y$), method.predict($X_{val}$)),. Thus, fitting all available methods does not require substantially more effort from the ML practitioner and will likely result in finding a best performing model. 

\begin{figure}[t]
 \centering
 \noindent\includegraphics[width=2.5in]{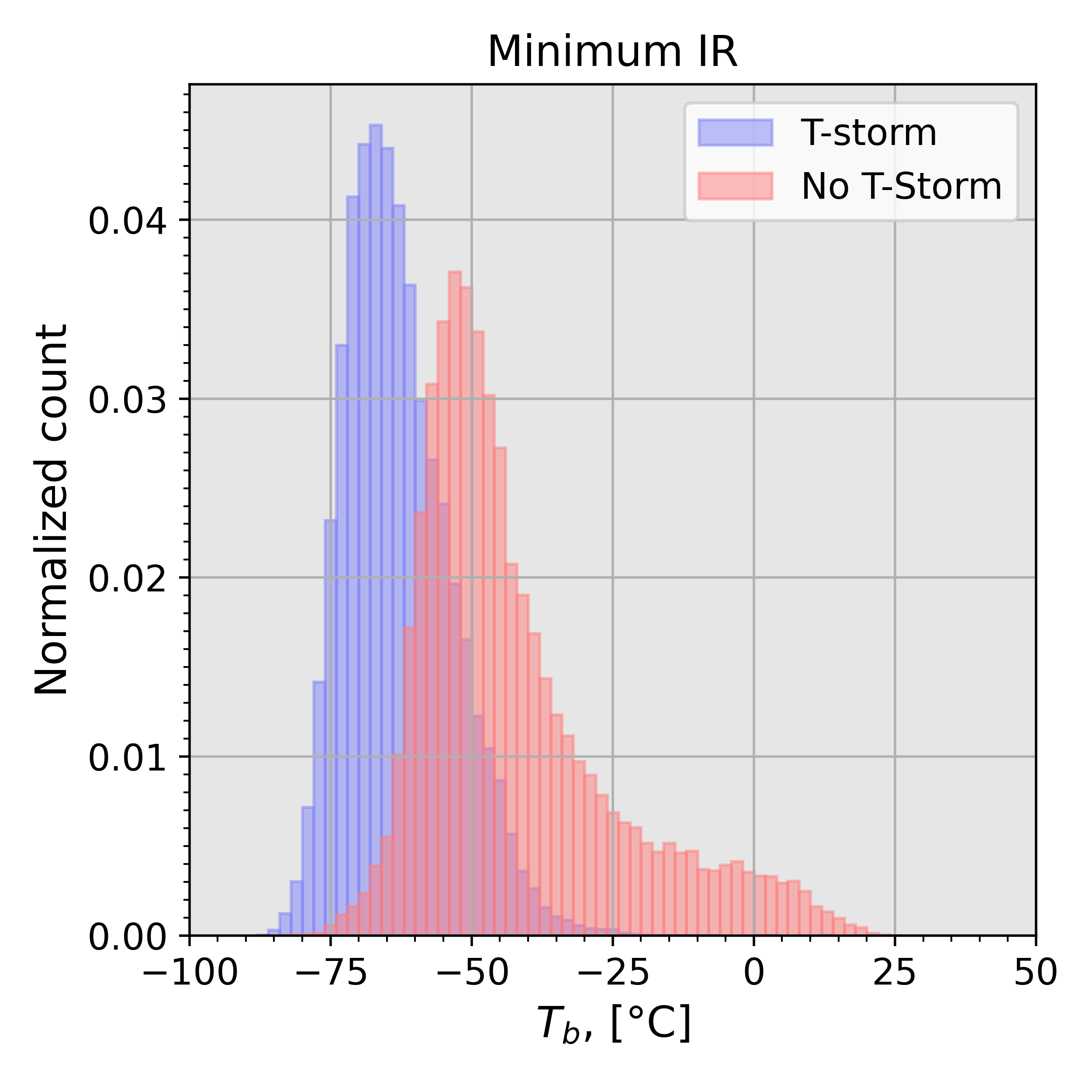}\\
 \caption{The normalized distributions of minimum brightness temperature ($T_{b}$) from the clean infrared channel for thunderstorm images (blue;T-storm) and non-thunderstorm images (red; No T-storm).}\label{Tb_fig_class}
\end{figure}

To start off, all methods are initially trained using their default hyperparameters in scikit-learn and just one input feature, the minimum infrared brightness temperature ($T_{b}$). We choose to use $T_{b}$ because meteorologically it is a proxy for the depth of the storms in the domain, which is correlated to lightning formation \citep{Yoshida2009}. To assess the predictive power of this variable, the distributions of $T_{b}$ for thunderstorms and no thunderstorms are shown in Fig. \ref{Tb_fig_class}. As expected, $T_{b}$ for thunderstorms show more frequent lower temperatures than non-thunderstorm images. Training all methods using $T_{b}$ achieves an accuracy of 80$\%$ on the validation dataset. While accuracy is a common and easy to understand metric, it is best to always use more than one metric when evaluating ML methods.

Another common performance metric for classification tasks is the Area Under the Curve (AUC). More specifically the common area metric is associated with the Receiver Operating Characteristics curve (ROC). The ROC curve is calculated from the relationship between the Probability of False Detection (POFD) and the Probability of Detection (POD). Both POFD and POD parameters are calculated from determining parameters within a contingency table which are the true positives (both the ML prediction and label say thunderstorm), false positives (ML prediction predicts thunderstorm, label has no thunderstorm), false negatives (ML prediction is no thunderstorm, label shows there is a thunderstorm) and true negatives (ML says no thunderstorm, label says no thunderstorm). The POFD and POD are defined by 

\begin{equation}
    \mathrm{POFD} = \frac{\mathrm{FalsePositive}}{\mathrm{TruePositive}+\mathrm{FalsePositive}}. \label{e15}
\end{equation}

\begin{equation}
    \mathrm{POD} = \frac{\mathrm{TruePositive}}{\mathrm{TruePositive}+\mathrm{FalseNegative}}. \label{e16}
\end{equation}

\begin{figure*}[t]
 \centering
 \noindent\includegraphics[width=6in]{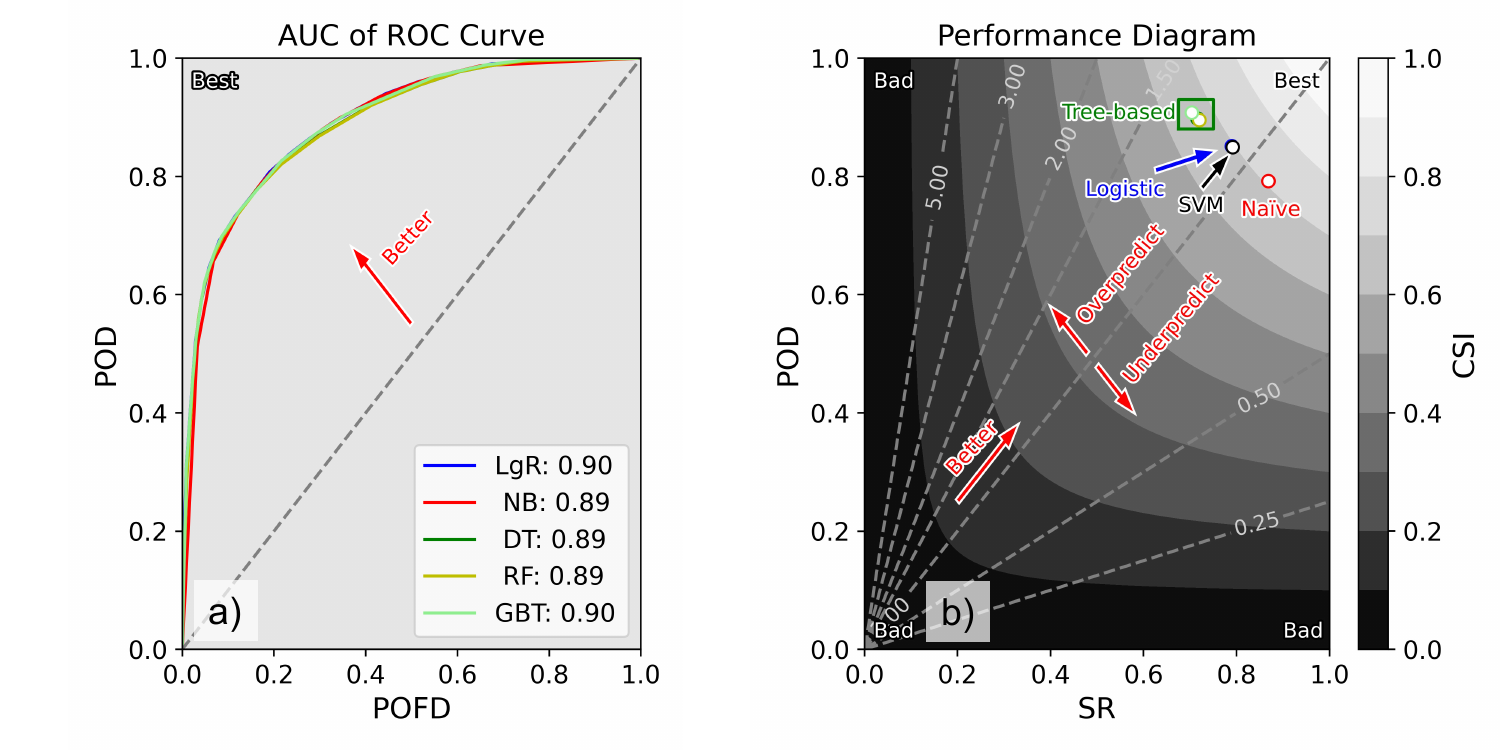}\\
 \caption{Performance metrics from the simple classification (only using $T_{b}$). (a) Receiver Operating Characteristic (ROC) curves for each ML model (except support vector machines), logistic regression (LgR; blue), naïve Bayes (NB; red), decision tree (DT; geen), random forest (RF; yellow) and gradient boosted trees (GBT; light green). The area under the ROC curve is reported in the legend. (b) Performance Diagram for all ML models (same colors as a). Color fill is the corresponding CSI value for each Success Ratio-Probability of Detection (SR, POD) pair. Dashed contours are the frequency bias.}\label{performance_class_simple}
\end{figure*}

All of the ML models, except support vector machines (as coded in sklearn), can provide a probabilistic estimation of the classification (e.g., this image is 95$\%$ likely to have lightning in it). When calculating the accuracy before, we assumed a threshold of 50$\%$ to designate what the ML prediction was. In order to get the ROC curve, the threshold probability is instead varied from 0$\%$ to 100$\%$. The resulting ROC curves for all of the ML methods except support vector machines are shown in Fig. \ref{performance_class_simple}a. We see that for this simple one feature model, all methods are still very similar and have AUCs near 0.9 (Fig. \ref{performance_class_simple}a), which is generally considered good performance\footnote{No formal peer reviewed journal states this, it is more of a rule of thumb in machine learning practice}. 

An additional method for evaluating the performance of classification method is called a performance diagram \citep[Figure \ref{performance_class_simple}b;][]{Roebber2009}. The performance diagram is also calculated from the contingency table, using the POD again for the y-axis, but this time the x-axis is the success ratio (SR) which is defined as
\begin{equation}
    \mathrm{SR} = \frac{\mathrm{TruePositive}}{\mathrm{TruePositive}+\mathrm{FalsePositive}}. \label{e17}
\end{equation}  
From this diagram, several things can be gleaned about the models performance. In general, the top right corner is where 'best' performing models are found. This area is characterized by models that capture nearly all events (i.e., thunderstorms), while not predicting a lot of false alarms (i.e., false positives). This corner is also associated with high values of critical success index (CSI; filled contours Fig. \ref{performance_class_simple}b), defined as
\begin{equation}
    \mathrm{CSI} = \frac{\mathrm{TruePositive}}{\mathrm{TruePositive}+\mathrm{FalsePositive}+\mathrm{FalseNegative}}. \label{e18}
\end{equation} 
which is a metric that shows a model's performance without considering the true negatives. Not considering the true negatives is important because true negatives can dominate ML tasks in meteorology given the often rare nature of events with large impacts (e.g., floods, hail, tornadoes). The last set of lines on this diagram are the frequency bias contours (dashed grey lines Fig. \ref{performance_class_simple}b). These contours indicate if a model is over-forecasting or under-forecasting. 

For the simple ML models trained, even though most of them have a similar accuracy and AUC, the performance diagram suggests their performance is indeed different. Consider the tree based methods (green box; Fig. \ref{performance_class_simple}b). They are all effectively at the same location with a POD of about 0.9 and a SR of about 0.75, which is a region that has a frequency bias of almost 1.5. Meanwhile the logistic regression, support vector machines and naïve Bayes methods are much closer to the frequency bias line of 1, while having a similar CSI as the tree based methods. Thus, after considering overall accuracy, AUC and the performance diagram, the best performing model would be either the logistic regression, support vector machines or naïve Bayes. At this junction, the practitioner has the option to consider if they want a slightly over-forecasting system or a slightly under-forecasting system. For the thunderstorm, no-thunderstorm task, there are not many implications for over-forecasting or under-forecasting. However, developers of a tornado prediction model may prefer a system that produces more false positives (over-forecasting; storm warned, no tornado) than false negatives (under-forecasting; storm not warned, tornado) as missed events could have significant impact to life and property. It should be clear that without going beyond a single metric, this differentiation between the ML methods would not be possible.

While the previous example was simple by design, we as humans could have used a simple threshold at the intersection of the two histograms in Fig. \ref{Tb_fig_class} to achieve similar accuracy (e.g., 81$\%$; not shown). The next logical step with the classification task would be to use all available features. One important thing to mention at this step is that it is good practice to normalize input features. Some of the ML methods (e.g., random forest) can handle inputs of different magnitudes (e.g., CAPE is on the order of 100s to 1000s, but Lifted Index is on the order of 1s to 10s), but others (e.g., logistic regression) will be unintentionally biased towards larger magnitude features if you do not scale your input features. Common scaling methods include min-max scaling and scaling your input features to have mean 0 and standard deviation of 1 (i.e., standard anomaly) which are defined mathematically as 
\begin{equation}
    \mathrm{minmax} = \frac{x - x_{min}}{x_{max} - x_{min}}. \label{e19}
\end{equation}

and 

\begin{equation}
    \mathrm{standard \ anom.} = \frac{x -\mu}{\sigma} \label{e20}
\end{equation}
respectively. In Eq. \ref{e19}, $x_{min}$ is the minimum value within the training dataset for some input feature $x$ while $x_{max}$ is the maximum value in the training dataset. In Eq. \ref{e20}, $\mu$ is the mean of feature $x$ in the training dataset and $\sigma$ is the standard deviation. For this paper, the standard anomaly is used. 

\begin{figure*}[t]
 \centering
 \noindent\includegraphics[width=6in]{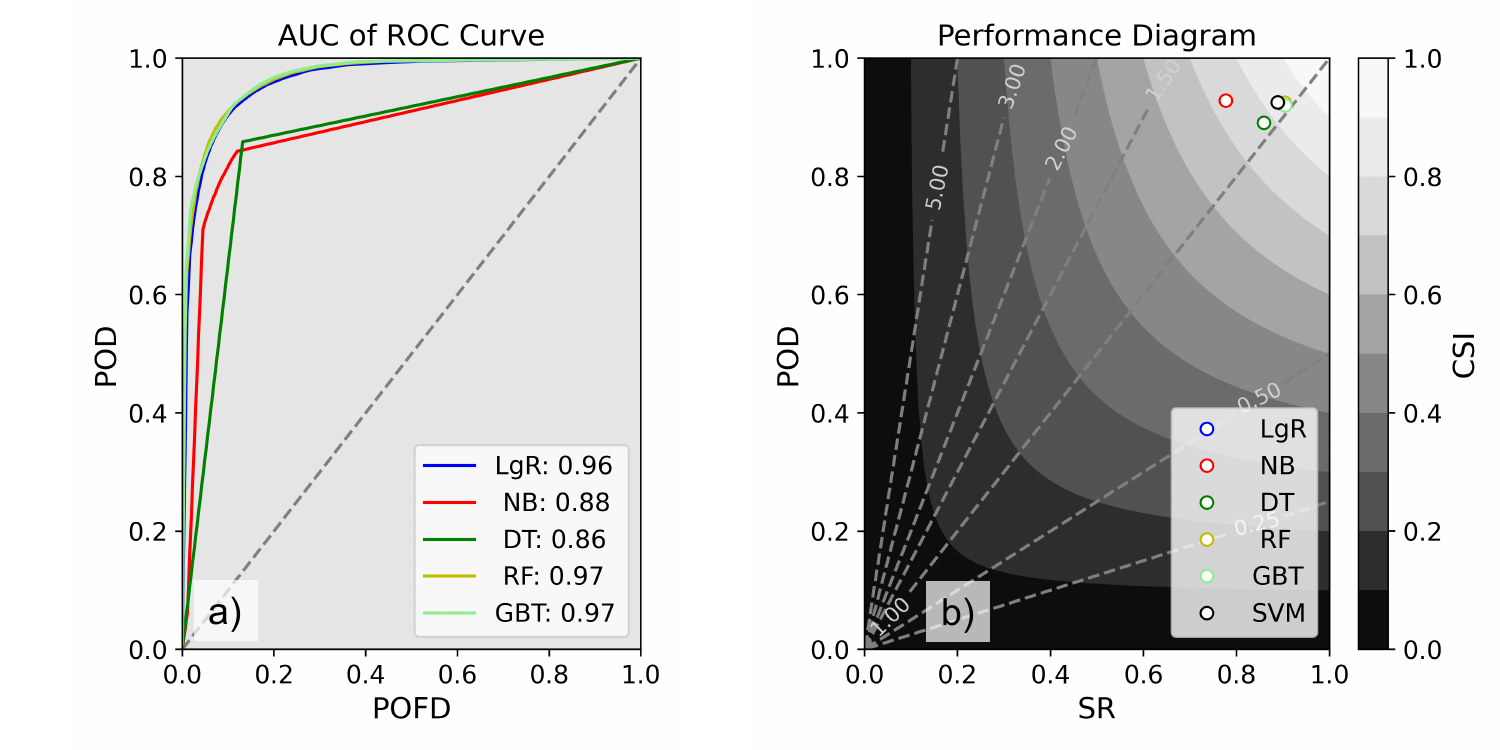}\\
 \caption{As in Figure \ref{performance_class_simple}, but now trained with all available predictors. The annotations from Fig. \ref{performance_class_simple} have been removed.} \label{performance_class_complex}
\end{figure*}

Using all available input features yields an accuracy of 90$\%$, 84$\%$, 86$\%$, 91$\%$, 90$\%$, 89$\%$ for logistic regression, naïve Bayes, decision tree, random forest, gradient boosted trees and support vector machines respectively. Beyond the relatively good accuracy, the ROC curves are shown in Fig. \ref{performance_class_complex}a. This time there are generally two sets of curves, one better performing group (logistic regression, random forest, gradient boosted trees and support vector machines) with AUCs of 0.97 and a worse performing group (naïve Bayes and decision tree) AUCs around 0.87. This separation coincides with the flexibility of the classification methods. The better performing groups are better set to deal with many features and non-linear interactions of the features, while the worse performing group is a bit more restricted in how it combines many features. Considering the performance diagram (Fig. \ref{performance_class_complex}b), the same grouping of high AUC performing models have higher CSI scores (> 0.8) and have little to no frequency bias. Meanwhile the lower AUC performing models have lower CSI (0.75) and NB has a slight overforecasting bias. Overall, the ML performance on classifying if an image has a thunderstorm is doing well with all predictors. While a good performing model is a desired outcome of ML, at this point we do not know how the ML is making its predictions. This is part of the 'black-box' issue of ML and does not lend itself to being consistent with the ML user's prior knowledge (see note in introduction on consistency; Murphy 1993). 

In order to alleviate some of opaqueness of the ML black-box, one can interrogate the trained ML models by asking: "What input features are most important to the decision?" and "Are the patterns the ML models learned physical (e.g., follow meteorological expectation)?". The techniques named permutation importance \citep[][]{Breiman2001, Lakshmanan2015} and accumulated local effects \citep[ALE;][]{Apley2020} are used to answer these two questions respectively. Permutation importance is a method in which the relative importance of a input feature is quantified by considering the change in evaluation metric (e.g., AUC) when that input variable is shuffled (i.e., randomized). The intuition is that the most important variables when shuffled will cause the largest change to the evaluation metric. There are two main flavors of permutation importance, named single-pass and multi-pass. Single-pass permutation importance goes through each input variable and shuffles them one by one, calculating the change in the evaluation metrics. Multi-pass permutation importance uses the result of the single-pass, but progressively permutes features. In other words, features are successively permuted in the order that they were determined as important (most important then second most important etc) from the single pass but are now left shuffled. The specific name for the method we have been describing is the \textit{backward multi-pass permutation importance}. The backward name comes from the direction of shuffling, starting will all variables unshuffled and shuffling more and more of them. There is the opposite direction, named \textit{forward multi-pass permutation importance}, where the starting point is that all features are shuffled to start. Then each feature is unshuffled in order of their importance from the single-pass permutation importance. For visual learners, see the animations (for the backward direction; Fig. ES4 and Fig. ES5) in the supplement of \citet{McGovern2019_bams}. The reason for doing multi-pass permutation importance is because correlated features could result in falsely identifying non-important variables using the single pass permutation importance. The best analysis of the permutation test is to use both the single pass and multi-pass tests in conjunction. 

\begin{figure*}[t]
 \centering
 \noindent\includegraphics[width=6in]{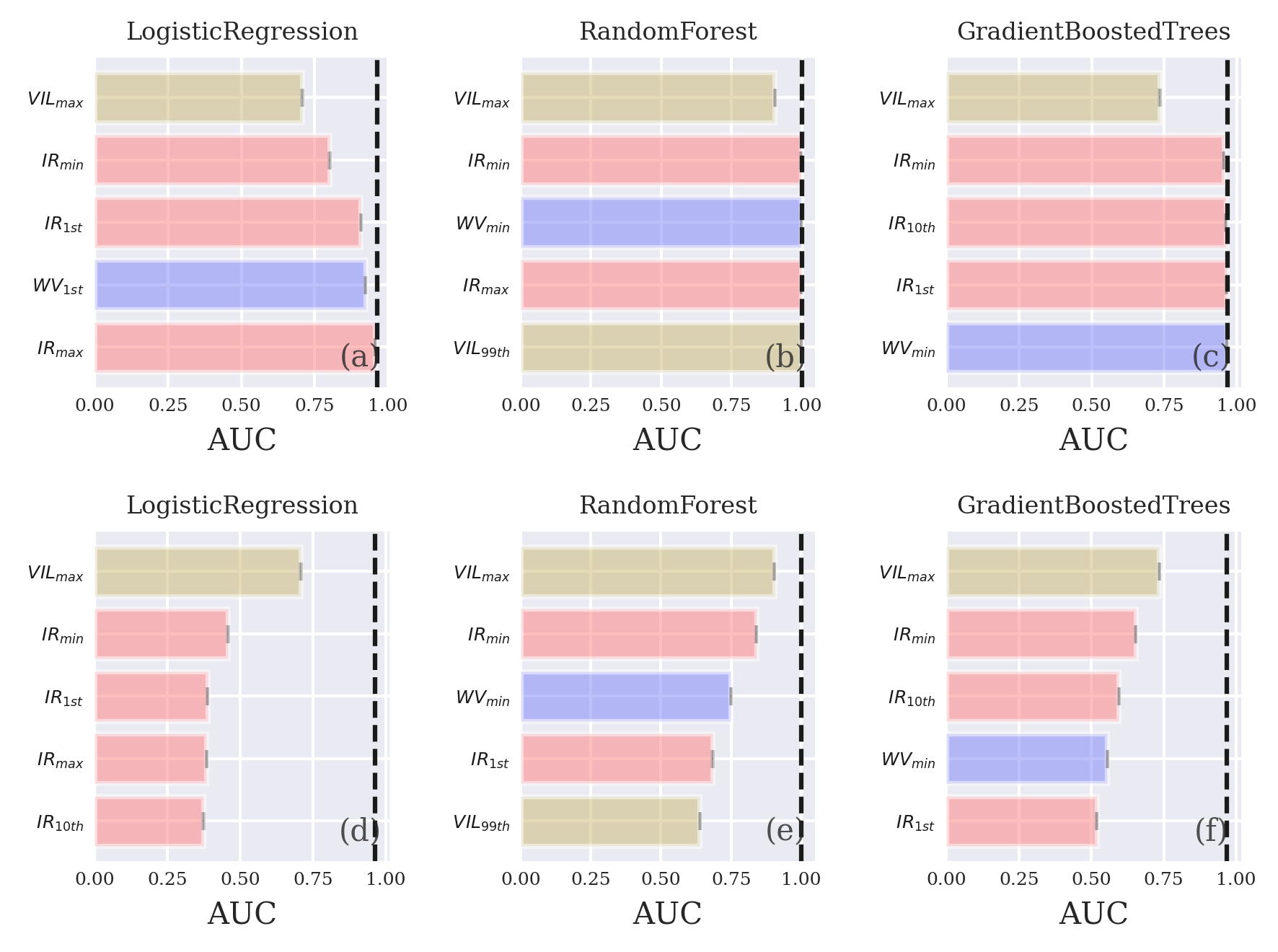}\\
 \caption{Backward permutation importance test for the best performing classification ML models. Single pass results are in the top row, while multi-pass forward results are for the bottom row. Each column corresponds to a different ML method: logistic regression (a,d), random forest (b,e) and gradient boosted trees (c,f). Bars are colored by their source, yellow for the vertically integrated liquid (VIL), red for the infrared (IR), blue for water vapor (WV) and black for visible (VIS). Number subscripts correspond to the percentile of that variable. The dashed black line is the original AUC value when all features are not shuffled.} \label{permutation_test}
\end{figure*}

The top five most important features for the better performing models (i.e., logistic regression, random forest and gradient boosted trees) as determined by permutation importance are shown in Fig. \ref{permutation_test}. For all ML methods both the single and multi-pass test show that the maximum vertically integrated liquid is the most important feature, while the minimum brightness temperature from the clean infrared and midtropospheric water vapor channels are found within the top 5 predictors (except multi-pass test for logistic regression). In general, the way to interpret these are to take the consensus over all models which features are important. At this point it time to consider if the most important predictors make meteorological sense. Vertically integrated liquid has been shown to have a relationship to lightning \citep[e.g.,][]{Watson1995} and is thus plausible to be the most important predictor. Similarly, the minimum brightness temperature at the water vapor and clean infrared channels also makes physical sense because lower temperatures are generally associated with taller storms. We could also reconcile the maximum infrared brightness temperature (Fig \ref{permutation_test}a) as a proxy for the surface temperature which correlates to buoyancy, but note that the relative change in AUC with this feature is quite small. Conversely, any important predictors that don't align with traditional meteorological knowledge may require further exploration to determine why the model is placing such weight on those variables. Does the predictor have some statistical correlation with the meteorological event that is unexplained by past literature, or are there nonphysical characteristics of the data that may be influencing the model during training? In the latter case, it is possible that your model might be getting the right answer for the wrong reasons.

Meanwhile minimum brightness temperature at both the water vapor and clean infrared channels also make physical sense since lower temperatures are related with taller storms. We could also reconcile the max infrared brightness temperature  as a proxy for the surface temperature, which correlates to buoyancy, but not that the relative change in AUC with this feature is quite small. If any the top predictors don't make sense meteorologically, then your model might be getting the right answer for the wrong reasons. 

\begin{figure*}[t]
 \centering
 \noindent\includegraphics[width=6in]{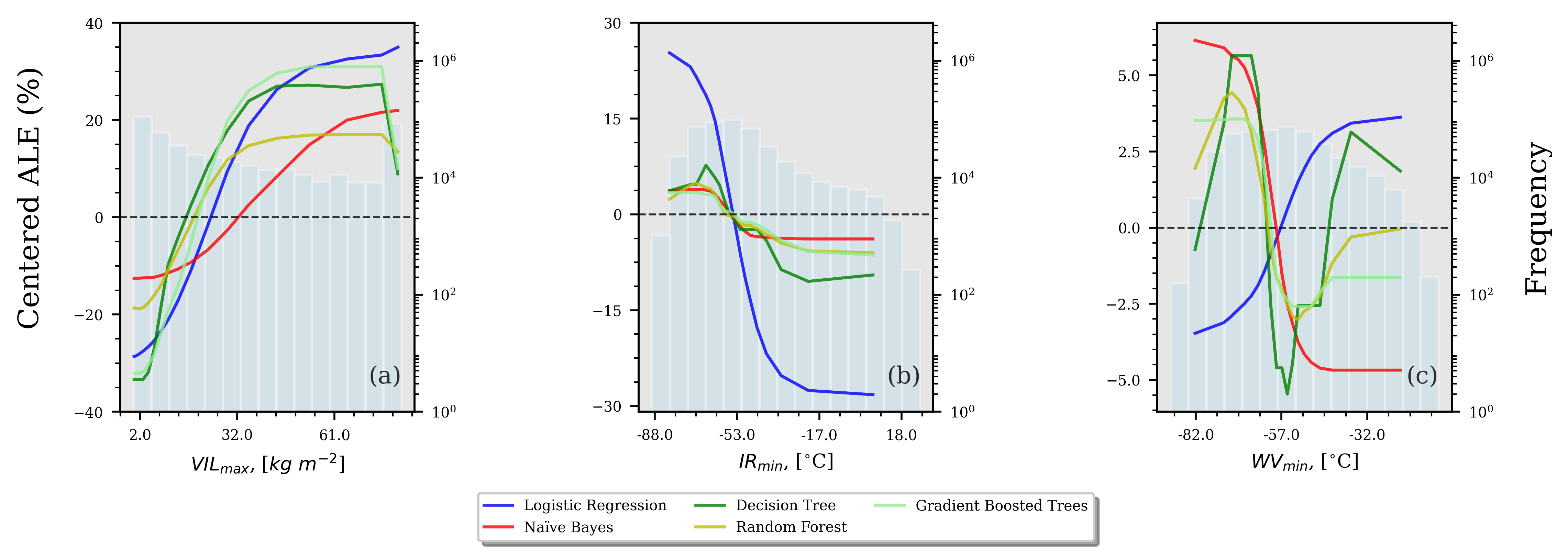}\\
 \caption{Accumulated local effects (ALE) for (a) the maximum Vertically Integrated Liquid ($\mathrm{VIL_{max}}$), (b) the minimum brightness temperature from infrared ($\mathrm{IR_{min}}$) and (c) the minimum brightness temperature from the water vapor channel ($\mathrm{WV_{min}}$). Lines correspond to all the ML methods trained (except support vector machines) and colors match Fig. \ref{performance_class_simple}. Grey histograms in the background are the counts of points in each bin.} \label{ale}
\end{figure*}

Accumulated local effects are where small changes to input features and their associated change on the output of the model are quantified. The goal behind ALE is to investigate the relationship between an input feature and the output. ALE is performed by binning the data based on the feature of interest. Then for each example in each bin, the feature value is replaced by the edges of the bin. The mean difference in the model output from the replaced feature value is then used as the ALE for that bin. This process is repeated for all bins which result in a curve. For example, the ALE for some of the top predictors of the permutation test are shown in in Fig. \ref{ale}. At this step, the ALEs can be mainly used to see if the ML models have learned physically plausible trends with input features. For the vertically integrated liquid, all models show that as the max vertically integrated liquid increases from about 2 $kg \ m^{2}$ to 30  $kg \ m^{2}$ the average output probability of the model will increase, but values larger than 30 $kg \ m^{2}$ generally all have the same local effect on the prediction (Fig. \ref{ale}a). As for the minimum clean infrared brightness temperature, the magnitude of the average change is considerably different across the different models, but generally all have the same pattern. As the minimum temperature increases from -88$^{\circ}$C to -55$^{\circ}$C, the mean output probability decreases: temperatures larger than -17 $^{\circ}$C have no change (Fig. \ref{ale}b). Lastly, all but the logistic regression shows a similar pattern with the minimum water vapor brightness temperature, but notice the magnitude of the y-axis (Fig. \ref{ale}c). Much less change occurs with this feature. For interested readers, additional interpretation techniques and examples can be found in \citet{Molnar2022}.

\subsubsection{Regression}

\begin{figure}[t]
 \centering
 \noindent\includegraphics[width=3in]{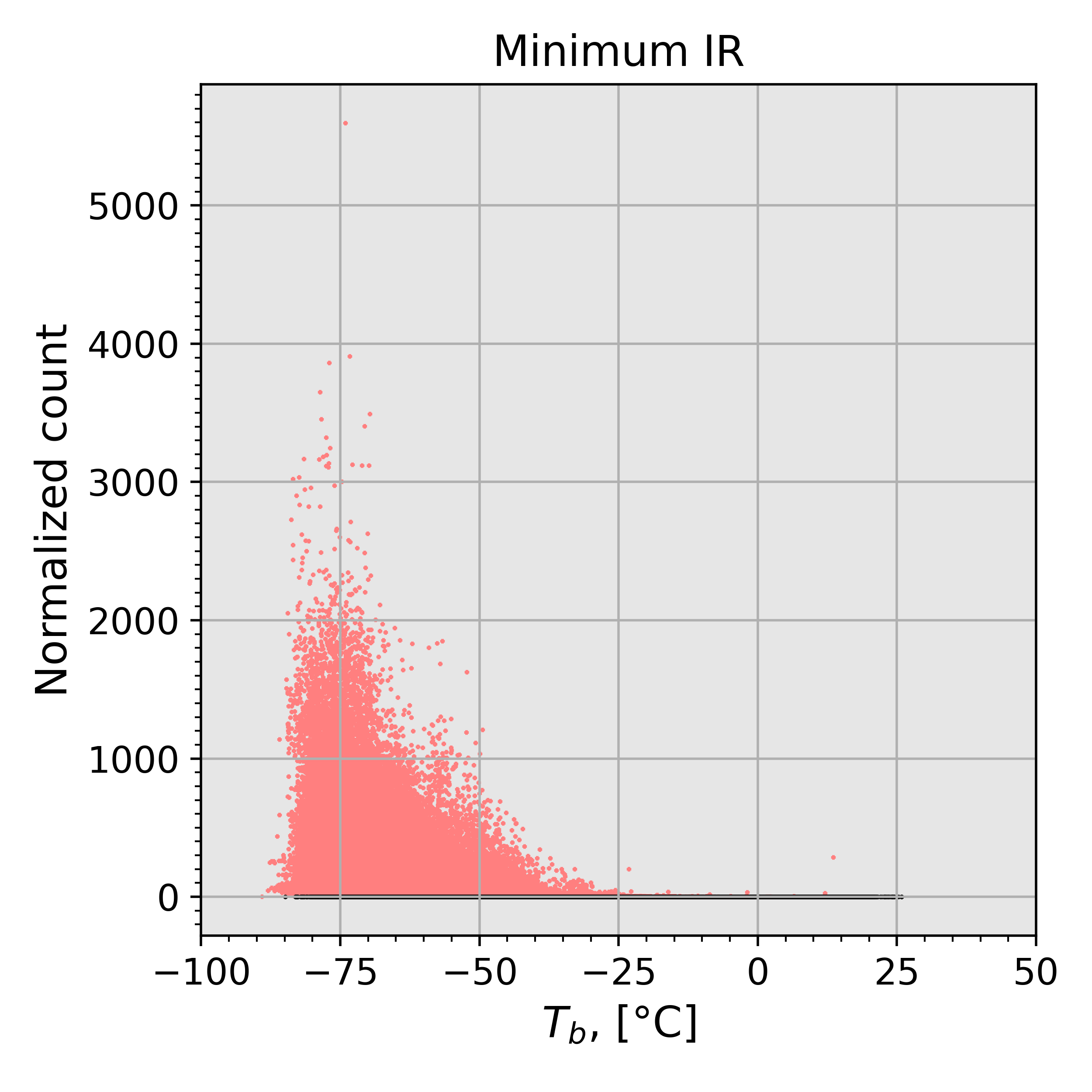}\\
 \caption{The training data relationship between the minimum brightness temperature from infrared ($T_b$) and the number of flashes detected by GLM. All non-thunderstorm images (number of flashes equal to 0) are in black.} \label{tb_flash}
\end{figure}

As stated in Section 3.a, task (2) is to predict the number of lightning flashes inside an image. Thus, the regression methods available to do this task are: linear regression, decision tree, random forest, gradient boosted trees and support vector machines. Similar to task (1) a simple scenario is considered first, using $T_{b}$ as the lone predictor. Figure \ref{tb_flash} shows the general relationship between $T_{b}$ and the number of flashes in the image. For $T_{b}$ > -25$^{\circ}C$, most images do not have any lightning, while $T_{b}$ < -25$^{\circ}C$ shows a general increase of lightning flashes. Given there are a lot of images with 0 flashes (approximately 50$\%$ of the total dataset; black points in Fig. \ref{tb_flash}), the linear methods will likely struggle to capture a skillful prediction. One way to improve performance would be to only predict the number of flashes on images where there is non-zero flashes. While this might not seem like a viable way forward since non-lightning cases would be useful to predict, in practice we could leverage the very good performance of the classification model from Section 3.c.1, and then use the trained regression on images that are confident to have at least one flash in them. An example of this done in the literature is \citet{Gagne2017_hail} where hail size predictions were only made if the classification model said there was hail. 

\begin{figure*}[t]
 \centering
 \noindent\includegraphics[width=6in]{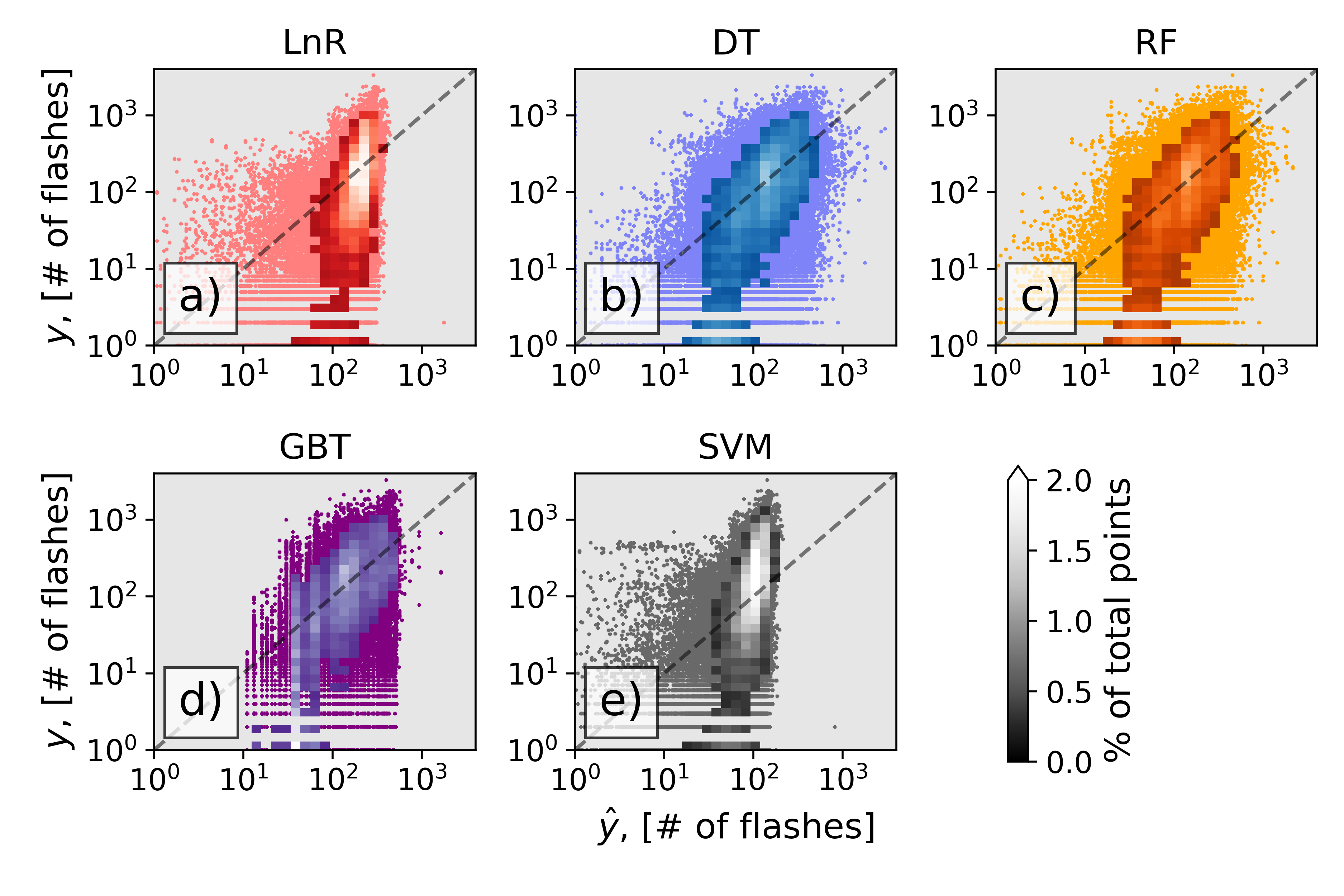}\\
 \caption{The one-to-one relationship between the predicted number of lightning flashes from the ML learning models trained on only $T_{b}$ (x-axis; $\hat{y}$) and the number of measure flashes from GLM (y-axis; $y$). Each marker is one observation. Meanwhile areas with more than 100 points in close proximity are shown in the colored boxes. The lighter the shade of the color, the higher density of points. (a) linear regression (LnR; reds), (b) decision tree (DT; blues), (c) random forest (RF; oranges), (d) gradient boosted trees (GBT; purples) and (e) linear support vector machines (SVM; greys).} \label{one_to_one_simple}
\end{figure*}

As before, all methods are fit on the training data initially using the default hyperparameters. A common way to compare regression model performance is to create a one-to-one plot, which has the predicted number of flashes on the x-axis and the true measured number of flashes on the y-axis. A perfect model will show all points tightly centered along the diagonal of the plot. This is often the quickest qualitative assessment of how a regression model is performing. While $T_{b}$ was well suited for the classification of thunderstorm/no-thunderstorm, it is clear that fitting a linear model to the data in Fig. \ref{tb_flash} did not do well (Fig. \ref{one_to_one_simple}a,e), leading to a strong over-prediction of the number of lightning flashes in an images with less than 100 flashes, while under-predicting the number of flashes for images with more than 100 flashes. The tree based methods tend to do better, but there is still a large amount of scatter and an over estimation of storms with less than 100 flashes. 

In order to tie quantitative metrics to the performance of each model the following are common metrics calculated: Mean Bias, Mean Absolute Error (MAE), Root Mean Squared Error (RMSE) and coefficient of determination ($R^2$). Their mathematical representations are the following: 

\begin{equation}
    \mathrm{Bias} =  \frac{1}{N} \sum_{j=1}^{N} (y_j - \hat{y}_j) , \label{e21}
\end{equation}

\begin{equation}
    \mathrm{MAE} = \frac{1}{N} \sum_{j=1}^{N} |y_j - \hat{y}_j|  \label{e22}
\end{equation}

\begin{equation}
    \mathrm{RMSE} = \sqrt{\frac{1}{N}  \sum_{j=1}^{N} (y_j - \hat{y}_j)^{2}}  \label{e23}
\end{equation}

\begin{equation}
    \mathrm{R^{2}} = 1 - \frac{\sum_{j=1}^{N} (y_j - \hat{y}_j)^{2}}{\sum_{j=1}^{N} (y_j - \bar{y})^{2}} \label{e24}
\end{equation}

\begin{figure}[t]
 \centering
 \noindent\includegraphics[width=3in]{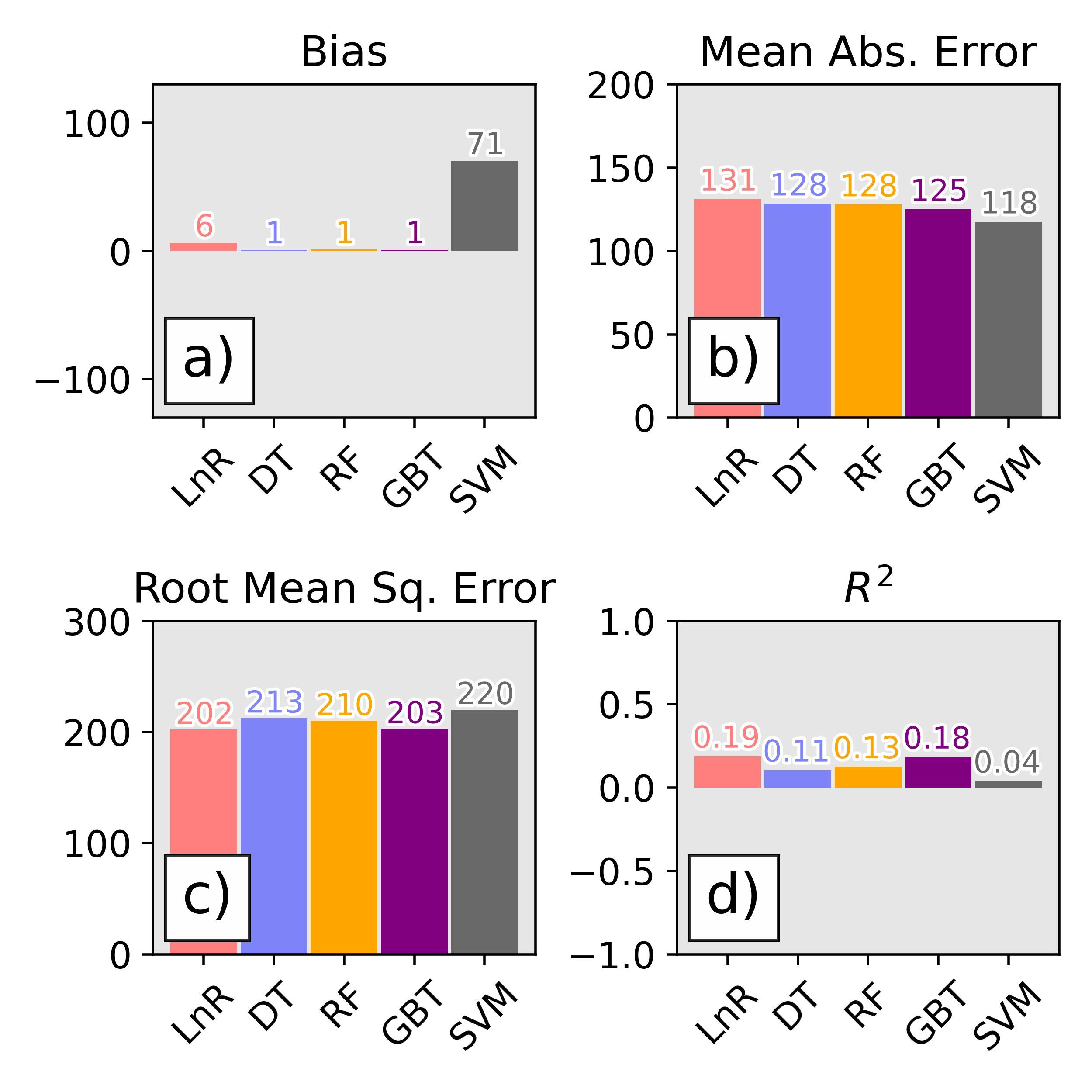}\\
 \caption{Validation dataset metrics for all ML models. Colors are the same as in Fig. \ref{one_to_one_simple}. Exact numerical value is reported on top of each bar.} \label{metrics_bar_simple}
\end{figure}

All of these metrics are shown in Fig. \ref{metrics_bar_simple}. In general, the metrics give a more quantitative perspective to the one-to-one plots. The poor performance of the linear methods shows, with the two worst performances being the support vector machines and linear regression with biases of 71 and 6 flashes respectively. While no method provides remarkable performance, the random forest and gradient boosted trees perform better with this single feature model (show better metrics holistically).

\begin{figure*}[t]
 \centering
 \noindent\includegraphics[width=6in]{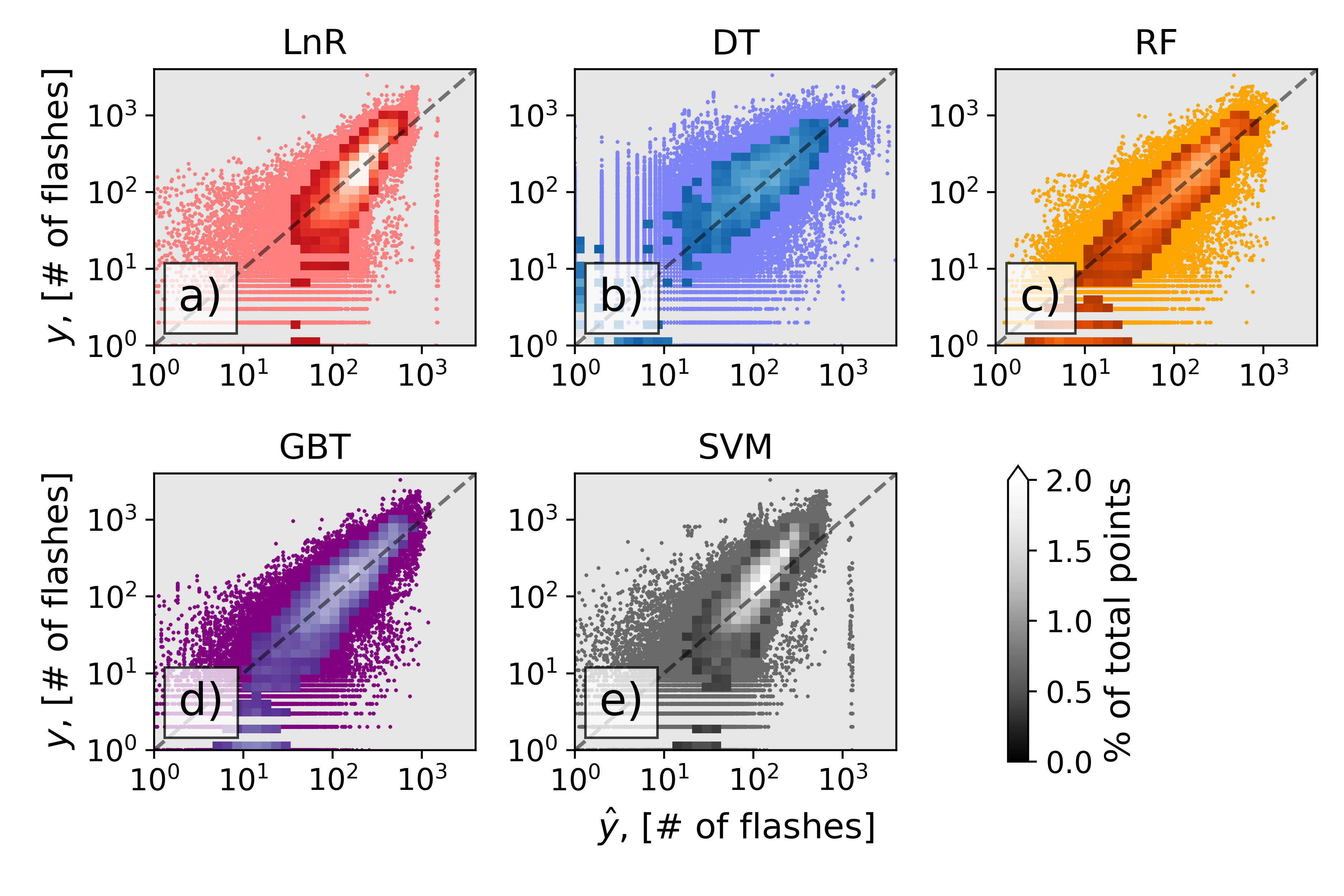}\\
 \caption{As in Fig. \ref{one_to_one_simple}, but now the x-axis is provided from the ML models trained with all available input features.} \label{one_to_one_complex}
\end{figure*}

\begin{figure}[t]
 \centering
 \noindent\includegraphics[width=3in]{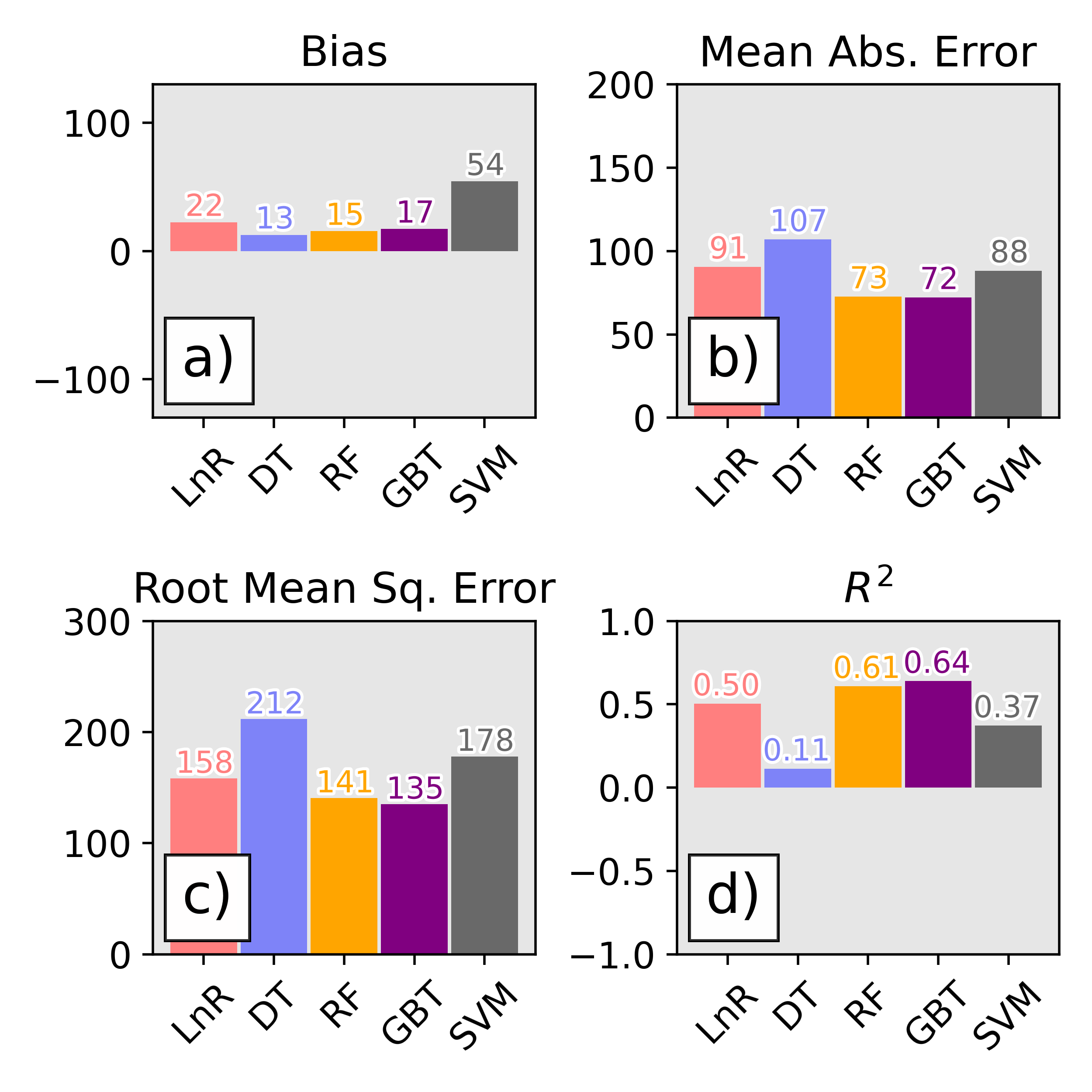}\\
 \caption{As in Fig. \ref{metrics_bar_simple}, but for ML models trained with all available input features.} \label{metrics_bar_complex}
\end{figure}

As before, the next logical step is to use all available features to predict the number of flashes: those results are found in Fig. \ref{one_to_one_complex} and \ref{metrics_bar_complex}. As expected, the model performance increases. Now all models show a general correspondence between the predicted number of flashes and the true number of flashes in the one-to-one plot (Fig. \ref{one_to_one_complex}). Meanwhile the scatter for random forest and gradient boosted trees has reduced considerably when comparing to the single input models (Fig. \ref{one_to_one_complex}c,d). While comparing the bias of the models trained with all predictors is relatively similar, the other metrics are much improved, showing large reductions in MAE, RMSE and increases in $R^{2}$ (Fig. \ref{metrics_bar_complex}) for all methods except decision trees. This reinforces that fact that similar to the classification example, it is always good to compare more than one metric. 

Since the initial fitting of the ML models used the default parameters, there might be room for tuning the models to have better performance. Here we will show an example of some hyperparameter tuning of a random forest. The common parameters that can be altered in a random forest include, the maximum depth of the trees (i.e., number of decisions in a tree) and the number of trees in the forest. The formal hyperparameter search will use the full training dataset, and systematically vary the depth of the trees from 1 to 10 (in increments of 1) as well as the number of trees from 1 to 100 (1,5,10,25,50,100). This results in 60 total models that are trained. 

\begin{figure}[t]
 \centering
 \noindent\includegraphics[width=3in]{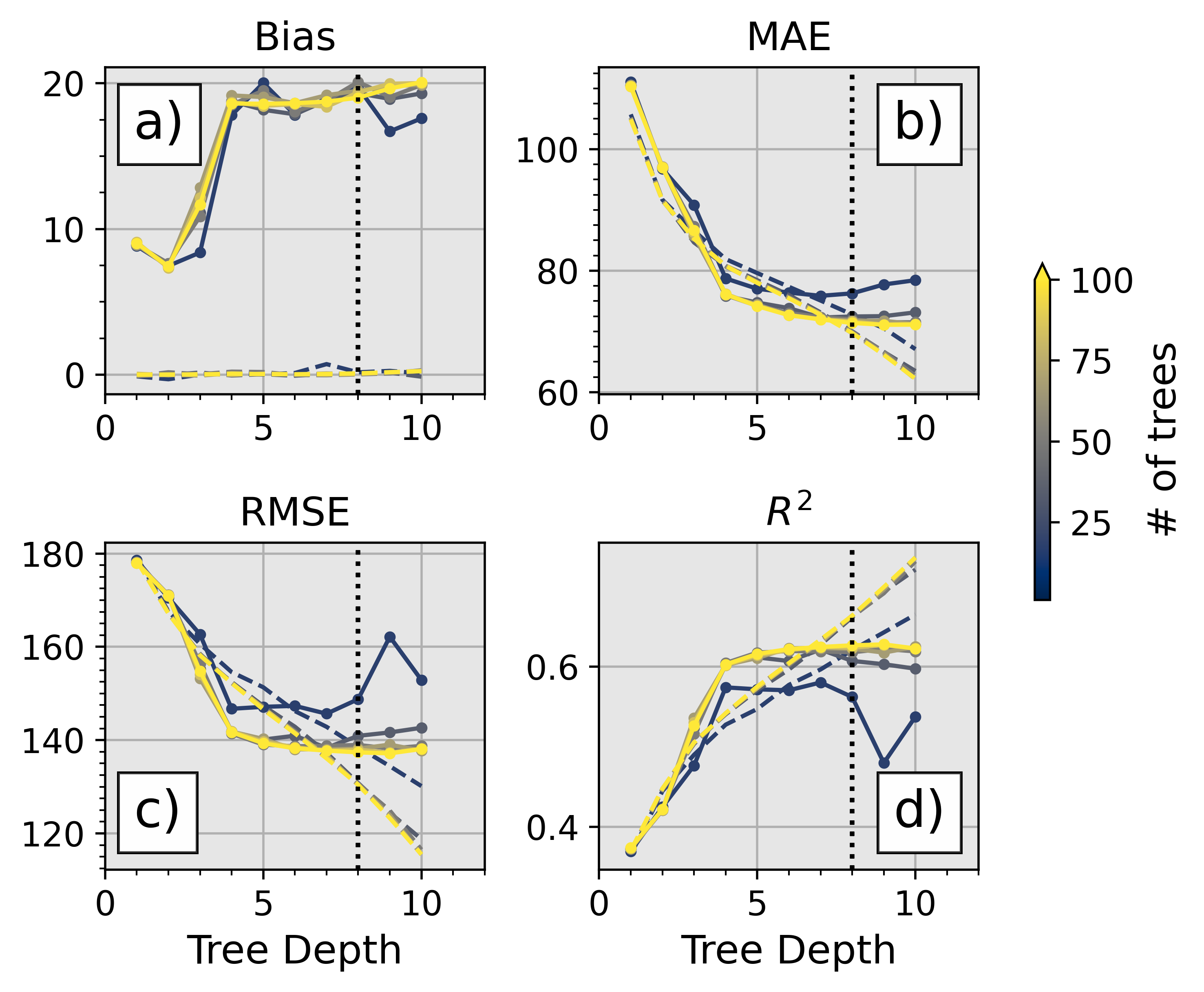}\\
 \caption{Hyperparameter tuning of a random forest for predicting the number of lightning flashes. All input features are used.Solid lines are the validation dataset while the dashed lines are the training data. The vertical dotted line is the depth of trees where over-fitting begins.} \label{hyper_tuning}
\end{figure}

In order to evaluate which is the best configuration, the same metrics as before are shown in Fig. \ref{hyper_tuning} as a function of the depth of the trees. The random forest quickly gains skill with added depth beyond one, with all metrics improving for both the training (dashed lines) and validation datasets (solid lines). Beyond a depth of four, the bias, MAE and RMSE all stagnate, but the $R^2$ value increases until a depth of eight where the training data continue to increase. There does not seem to be that large of an effect of increasing the number of trees beyond 10 (color change of lines). The characteristic of increasing training metric skills but no increase (or a decrease) to validation data skill is the over-fitting signal we discussed in Section 3.b. Thus, the best random forest model choice for predicting lightning flashes is a random forest with a max depth of eight and a total of 10 trees. The reason we choose 10 trees, is because in general choosing a simpler model is less computationally expensive to use as well as a more interpretable than a model with 1000 trees.

\subsection{Testing \label{sec:testing}}
As mentioned before, the test dataset is the dataset you hold out until the end when all hyperparameter tuning has finished so that there is no unintentional tuning of the final model configuration to a dataset. Thus, now that we have evaluated the performance of all our models on the validation dataset it is time to run the same evaluations as in Section 3.c.1 and Section 3.c.2. These test results are the end performance metrics that should be interpreted as the expected ML performance on new data (e.g., the ML applied in practice). For the ML models here the metrics are very similar as the validation set. For brevity the extra figures are included in the appendix (Fig A1-A3). 

\section{Summary and Future Work \label{subsec:Summary}}
This manuscript was the first of two Machine Learning (ML) tutorial papers designed for the operational meteorology community. This paper supplied a survey of some of the most common ML methods. All ML methods described here are considered \textit{supervised} methods, meaning the data the models are trained from include pre-labeled \textit{truth} data. The specific methods covered included linear regression, logistic regression, decision trees, random forests, gradient boosted decision trees, naïve Bayes and support vector machines. The overarching goal of the paper was to introduce the ML methods in such a way that ML methods are more familiar to readers as they encounter them in the operational community and within the general meteorological literature. Moreover, this manuscript provided ample references of published meteorological examples as well as open-source code to act as catalysts for readers to adapt and try ML on their own datasets and in their workflows. 

Additionally, this manuscript provided a tutorial example of how to apply ML to a couple meteorological tasks using the Storm EVent ImageRy dataset \citep[SEVIR;][]{Veillett2020} dataset. We: 

\begin{enumerate}
    \item{Discussed the various steps of preparing data for ML (i.e., removing artifacts; engineering features, train/val/test splits; Section 3.b)}
    \item{Conducted a classification task to predict if satellite images had lightning within them. This section included discussions of training, evaluation and interrogation of the trained ML models (Section 3.c.1)}
    \item{Exhibited a regression task to predict the number of lightning flashes in a satellite image. This section also contained discussions of training/evaluation as well as an example of hyperparameter tuning (Section 3.c.2)}
    \item{Released python code to conduct all steps and examples in this manuscript (see \textit{Data Availability Statement})}
\end{enumerate}

The follow on paper in this series will discuss a more complex, yet potentially more powerful, grouping of ML methods: neural networks and deep learning. Like a lot of the ML methods described in this paper, neural networks aren't necessarily new \citep{Rumelhart1986} and were first applied to meteorology topics decades ago \citep[e.g.,][]{Key1989,Lee1990}. Although, given the exponential growth of computing resources and dataset sizes, research using neural networks and deep learning in meteorology has been accelerating \citep[e.g., Fig 1c; ][]{Gagne_2019_gan,Lagerquist2020,Cintineo2020,Chase2021,Hilburn2021,Lagerquist2021,Molina2021,Ravuri2021}. Thus, it is important that operational meteorologists also understand the basics of neural networks and deep learning.

\clearpage
\acknowledgments
This material is based upon work supported by the National Science Foundation under Grant No. ICER-2019758, supporting authors RJC, AM and AB. Author DRH was provided support by NOAA/Office of Oceanic and Atmospheric Research under NOAA-University of Oklahoma Cooperative Agreements number NA16OAR4320115 and number NA21OAR4320204, U.S. Department of Commerce. The scientific results and conclusions, as well as any views or opinions expressed herein, are those of the authors and do not necessarily reflect the views of NOAA or the Department of Commerce.

We want to acknowledge the work put forth by the authors of the SEVIR dataset (Mark S. Veillette, Siddharth Samsi and Christopher J. Mattioli) for making a high-quality free dataset. We would also like to acknowledge the open-source python community for providing their tools for free. Specifically, we acknowledge Google Colab \citep{Bisong2019}, Anaconda \citep{anaconda}, scikit-learn \citep{scikit-learn}, Pandas \citep{mckinney-proc-scipy-2010}, Numpy \citep{harris2020array} and Jupyter \citep{Kluyver2016jupyter}. 


%
%
\datastatement
As an effort to catalyse the use and trust of machine learning within meteorology we have supplied a github repository with a code tutorial of a lot of the same things discussed in this paper. The latest version of github repository can be located here: \url{https://github.com/ai2es/WAF_ML_Tutorial_Part1}. If you are interested in the version of the repository that was available at time of publication please see the zendo archive of version 1 here: URL. The original github repo for SEVIR is located here: \url{https://github.com/MIT-AI-Accelerator/neurips-2020-sevir}. 


%

\appendix




\appendixtitle{Testing dataset figures}

\begin{figure}[t]
 \centering
 \noindent\includegraphics[width=6in]{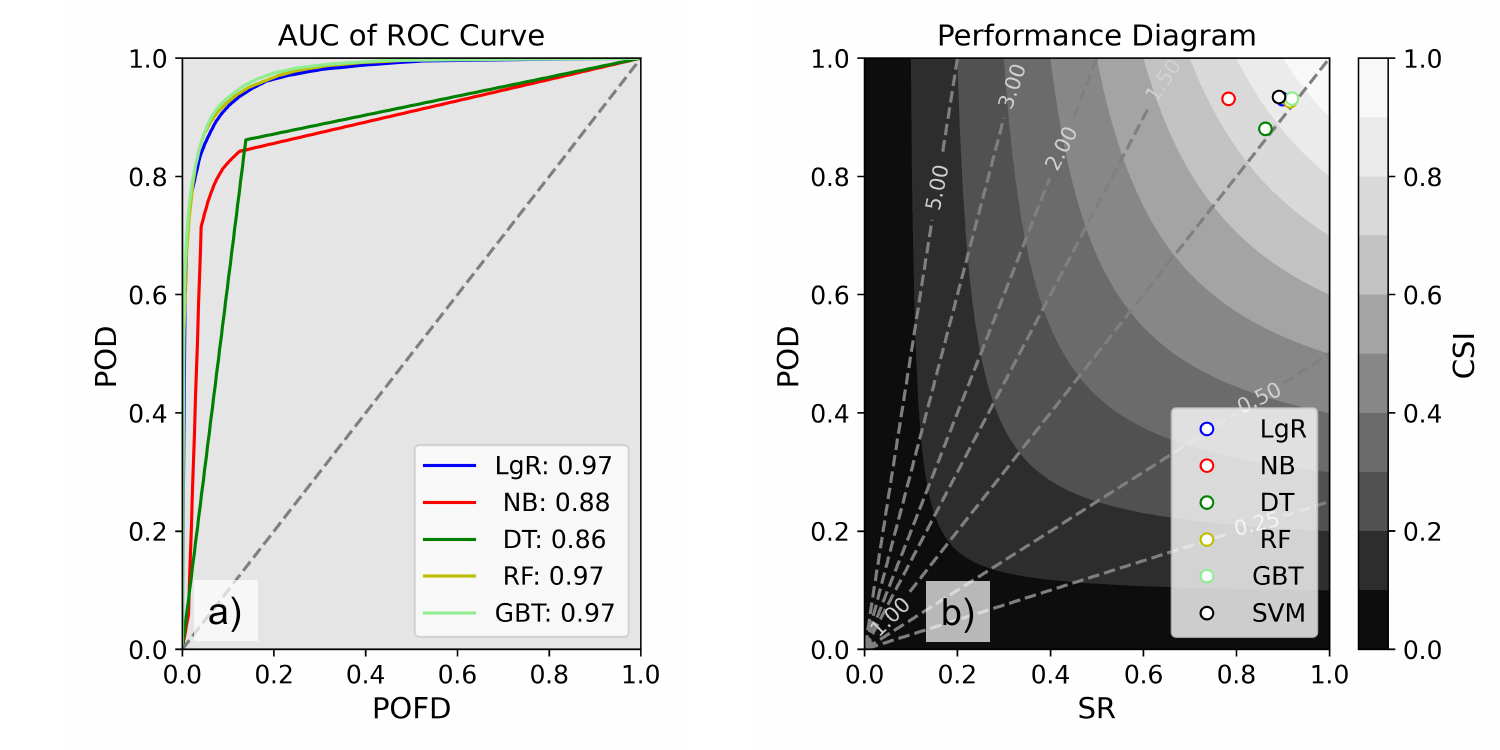}\\
 \caption{As in Figure \ref{performance_class_simple}, but now for the test dataset} \label{A1}
\end{figure}

\begin{figure}[t]
 \centering
 \noindent\includegraphics[width=6in]{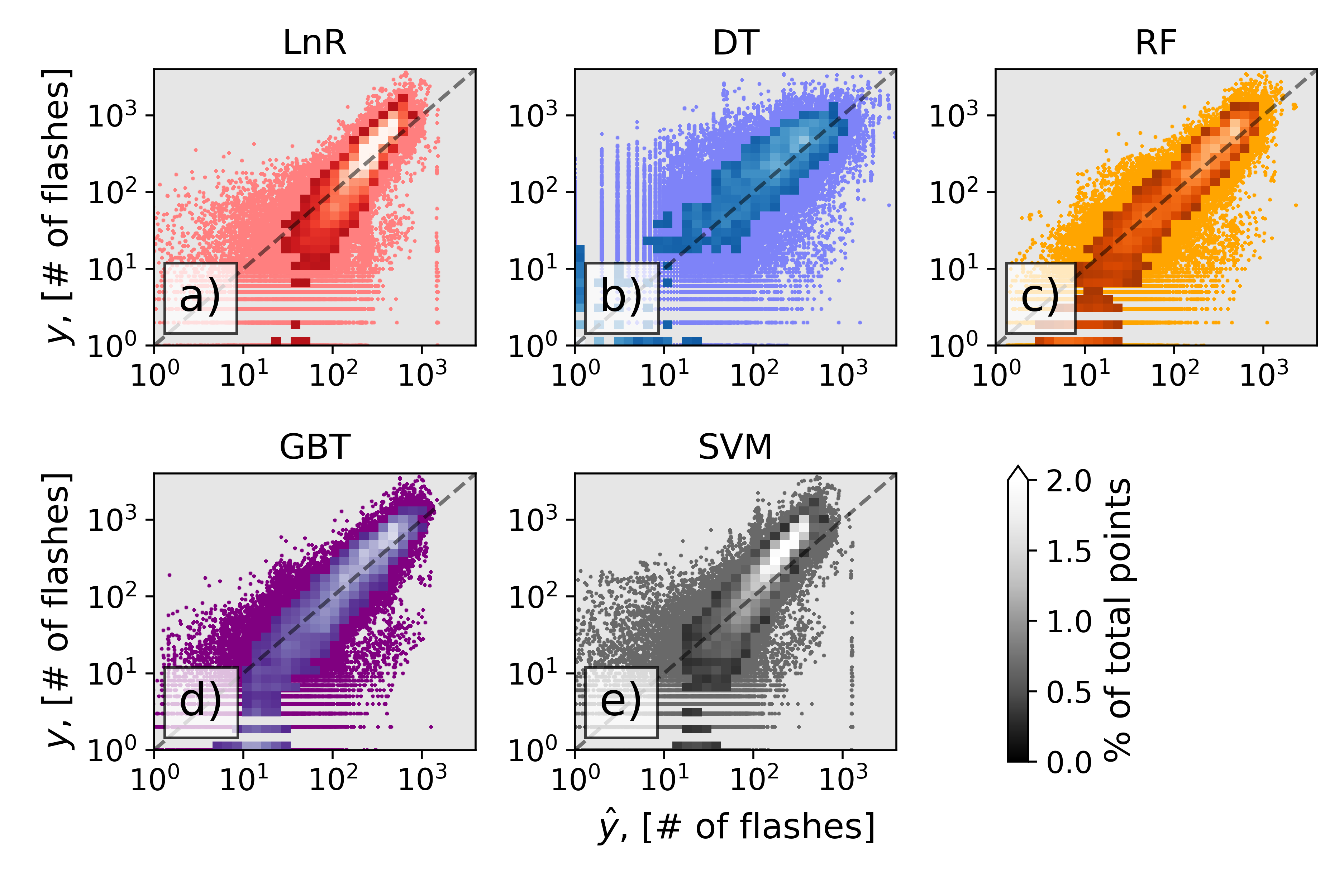}\\
 \caption{As in Fig. \ref{one_to_one_simple}, but for the test dataset} \label{A2}
\end{figure}

\begin{figure}[t]
 \centering
 \noindent\includegraphics[width=3in]{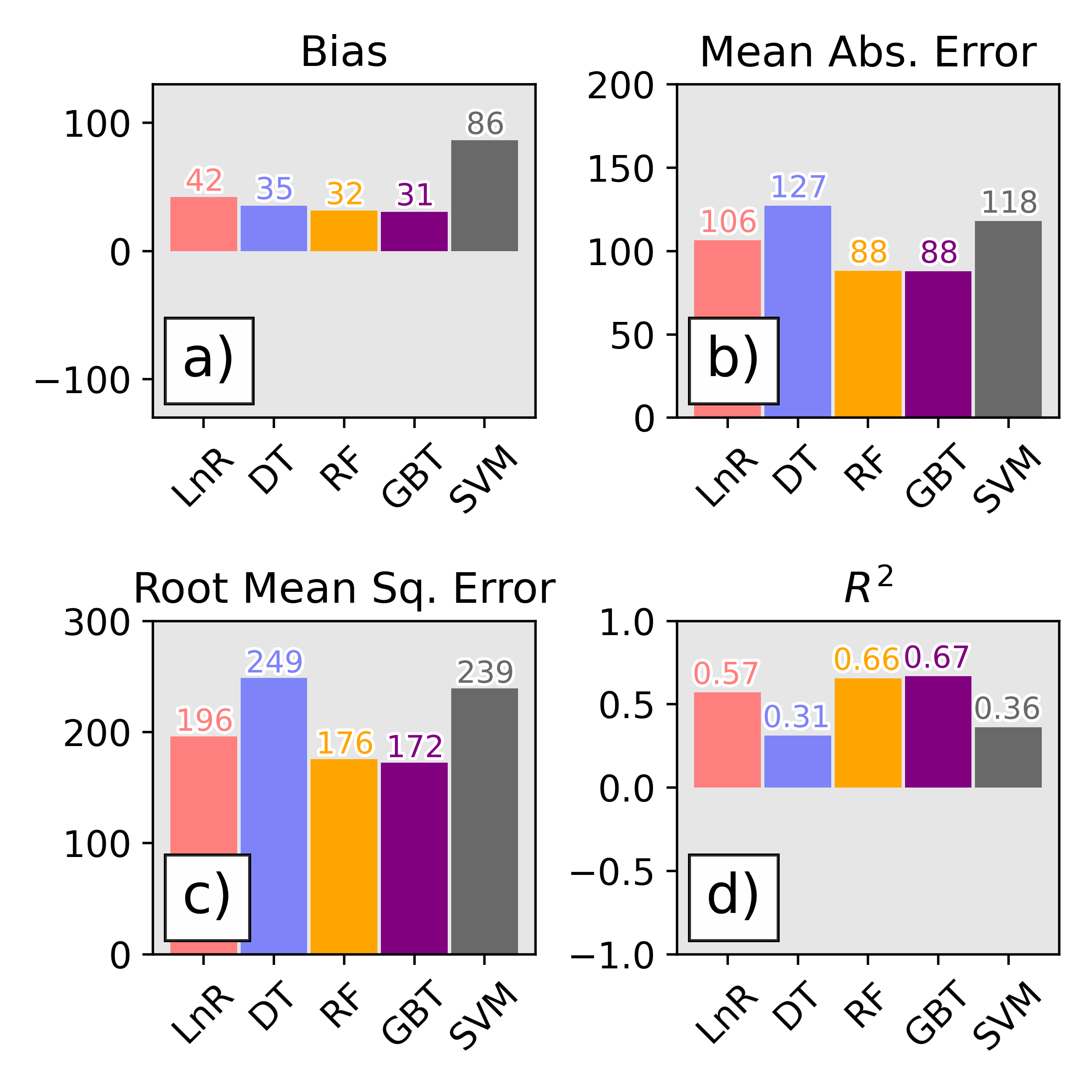}\\
 \caption{As in Fig. \ref{metrics_bar_simple}, but for the test dataset} \label{A3}
\end{figure}

This appendix contains the test dataset evaluations for both the classification task (Fig. \ref{A1}) and the regression task (Fig. \ref{A2}-\ref{A3}). Results are largely the same as the validation set, so to save space they were included here. 

\clearpage
\clearpage

\clearpage
\clearpage

\clearpage
\clearpage

%



\bibliographystyle{ametsocV6}
\bibliography{clean_references}

\end{document}